\documentclass[fleqn,usenatbib]{mnras}

\usepackage{newtxtext,newtxmath}

\usepackage[T1]{fontenc}
\usepackage{ae,aecompl}

\usepackage{graphicx}	
\usepackage{amsmath}	
\usepackage{amssymb}	
\usepackage{overpic}
\usepackage{natbib}

\title[GW170817 in X-rays]{A long-lived neutron star merger remnant in GW170817: constraints and  clues from X-ray observations 
}
\author[Piro et al.]{
L.~Piro$^{1}$ \thanks{E-mail: luigi.piro@iaps.inaf.it}
E.~Troja$^{2,3}$, 
B.~Zhang$^{4}$, 
G.~Ryan$^{2,12}$, 
H.~van~Eerten$^{5}$, 
R.~Ricci$^{6}$, 
\newauthor
M.~H.~Wieringa$^{7}$, 
A.~Tiengo$^{8,9,10}$, 
N.~R.~Butler$^{11}$,
S.~B.~Cenko$^{3,12}$, 
O.~D.~Fox$^{13}$,
\newauthor
H.~G.~Khandrika$^{13}$, 
G.~Novara$^{8,9}$,
A.~Rossi$^{14}$,
T.~Sakamoto$^{15}$
\\
$^{1}$ INAF, Istituto di Astrofisica e Planetologia Spaziali, via Fosso del Cavaliere 100, 00133 Rome, Italy \\
$^{2}$ Department of Astronomy, University of Maryland, College Park, MD 20742-4111, USA \\
$^{3}$ Astrophysics Science Division, NASA Goddard Space Flight Center, 8800 Greenbelt Rd, Greenbelt, MD 20771, USA \\
$^{4}$ Department of Physics and Astronomy, University of Nevada, 89154, Las Vegas, NV, USA \\
$^{5}$ Department of Physics, University of Bath, Claverton Down, Bath, BA2 7AY, UK \\
$^{6}$ INAF-Istituto di Radioastronomia, Via Gobetti 101, I-40129, Bologna, Italy \\
$^{7}$ CSIRO Astronomy and Space Science, PO Box 76, Epping, New South Wales 1710, Australia \\
$^{8}$ Scuola Universitaria Superiore IUSS Pavia, Piazza della Vittoria 15, 27100 Pavia, Italy \\
$^{9}$ INAF - IASF Milano, Via E. Bassini 15, 20133 Milano, Italy \\
$^{10}$ Istituto Nazionale di Fisica Nucleare, Sezione di Pavia, Via Bassi 6,
27100 Pavia, Italy \\
$^{11}$ School of Earth \& Space Exploration, Arizona State University, AZ 85287, USA \\
$^{12}$ Joint Space-Science Institute, University of Maryland, College Park, Maryland 20742, USA \\
$^{13}$ Space Telescope Science Institute, 3700 San Martin Drive, Baltimore, MD 21218, USA \\
$^{14}$INAF - OAS Bologna, Via P. Gobetti 93/3, 40129 Bologna, Italy\\
$^{15}$ Department of Physics and Mathematics, Aoyama Gakuin University, 5-10-1 Fuchinobe, Chuo-ku, Sagamihara-shi Kanagawa 252-5258, Japan
}

\date{Accepted XXX. Received YYY; in original form ZZZ}

\pubyear{2018}
\begin{document}
\label{firstpage}
\pagerange{\pageref{firstpage}--\pageref{lastpage}}
\maketitle

\begin{abstract}
Multi-messenger observations of GW170817 have not conclusively established whether the merger remnant is a black hole (BH) or a neutron star (NS).
We show that  a long-lived magnetized NS with a poloidal field $B\approx 10^{12}$G is fully consistent with the  electromagnetic dataset, when  spin down losses are dominated by gravitational wave (GW) emission.  The required  ellipticity $\epsilon\gtrsim 10^{-5}$ can  result from a  toroidal magnetic field component  much stronger than the poloidal component, a configuration expected from a NS newly formed from a merger.  Abrupt magnetic dissipation of the toroidal component can lead to the appearance of X-ray flares, analogous to the one observed in gamma-ray burst (GRB) afterglows.  	
In the X-ray afterglow of GW170817 we identify 
a low-significance  ($\gtrsim 3\sigma$) temporal feature at 155 d, consistent with a sudden reactivation of the central NS. Energy injection from the NS spin down into the relativistic shock is negligible, and the underlying continuum is fully accounted for by a structured jet seen off-axis. 
Whereas radio and optical observations probe the interaction of this jet with the surrounding medium,  observations at X-ray wavelengths, performed with adequate sampling,  open a privileged window on to the merger remnant. 
\end{abstract}
\begin{keywords}
gravitational waves -- gamma-ray burst: general -- neutron stars
\end{keywords}



\begin{table*}[!b]
 	\caption{X-ray observations of GW170817. Errors are 1~$\sigma$.}
 	\label{tab:xray}
 	\begin{tabular}{cccccc}
    	\hline
        $T-T_0$ & Exposure  & Count rate & Unabsorbed Flux  &  Flux density & Facility \\
        (d) & (ks) &  (10$^{-3}$ cts s$^{-1}$) & (10$^{-14}$ erg cm$^{-2}$ s$^{-1}$) &  (10$^{-3}$ $\mu$Jy)  \\
         & & 0.5--8.0 keV & 0.3 -- 10 keV & 1 keV & \\
        \hline
       153   &  32.1   & 2.0\,$\pm$\, 0.3 & 3.2\,$\pm$\,0.4  &  2.7\,$\pm$\,0.3  & {\it Chandra} \\ 
       157  &  16.0  &  2.0\,$\pm$\, 0.4 & 3.2\,$\pm$\,0.6  &  2.7\,$\pm$\,0.5  &" \\
       160  &  21.0 &  1.5\,$\pm$\, 0.3 & 2.6\,$\pm$\,0.5  &   2.2\,$\pm$\,0.4  & "\\      
       161 &  22.5  &  1.1\,$\pm$\, 0.3 & 1.8\,$\pm$\,0.4   & 1.5\,$\pm$\,0.3  & "\\
       163 &  110  &  1.36\,$\pm$\, 0.11  & 1.9\,$\pm$\,0.2 &    1.63\,$\pm$\,0.17  &  {\it XMM-Newton}\\
      165 &  14.4  &   1.0\,$\pm$\, 0.4 & 1.9\,$\pm$\,0.5 &  1.6\,$\pm$\,0.4  &  {\it Chandra}\\
       260 &  96.7  &   0.86\,$\pm$\, 0.17 & 1.2\,$\pm$\,0.2&   1.03\,$\pm$\,0.17  & "\\
      \hline
    \end{tabular}
\end{table*}

\section{Introduction}
 Pairs of neutron stars (NSs)
are bound to spiral into each other due to their  persistent emission of gravitational waves (GWs). 
Depending on the total mass of the system and the neutron star equation of state (EoS), the final product of the NS-NS merger can be either a black hole (BH) or a NS. 
Multi-messenger observations of GW170817, the first NS-NS merger system detected by advanced LIGO and advanced Virgo \citep{GW170817-multi}, have shown general consistency with a BH merger product, even though the possibility of a long-lived NS is not ruled out \citep{LIGO-product,ai18}. Indeed, the NS scenario has interesting implications on the kilonova (KN) models \citep{kasen15,gao15,radice18}, alleviating  demanding requirements on the mass of ejecta \citep{yu17,li18,metzger18}.
On the other hand, the radiation emitted from such long-lived NS 
should not violate the limits posed by the multi-wavelength observations of the GW counterpart \citep[e.g.][]{SwiftGW,margalit17,ai18,pooley18}. 

A common - although not unique - interpretation is  that the luminous blue component of the KN AT2017gfo was produced by lanthanide-poor accretion disc outflows along the binary polar axis \citep[e.g.][]{SwiftGW, Troja2017,kasen17,tanvir17,Smartt17,Pian17}
which is generally thought to produce less massive outflows, and support the immediate formation of a NS. 
\citet{margalit17} further constrained the nature of the relic NS  by correlating the observed GW and GRB emission. 
Growing observational evidence shows that the merger remnant launched a relativistic jet
\citep{Mooley18b,Troja18b,Ghirlanda18}, which powered the observed GRB and broadband afterglow emission.
In the standard GRB model the jet  is formed and launched by an accreting solar-mass BH, and the 1.7 s delay between the GW and GRB emission \citep{goldstein17,Savchenko17} could be interpreted as the maximum lifetime of the remnant NS \citep{metzger18}, after which it collapsed into a BH. 
Such short lifetime would favor the formation of a hypermassive NS (HMNS) \citep[e.g.][]{margalit17}.  
However, if the central NS was longer lived \citep{vanputten18} and launched the GRB outflow, different outcomes, such as a supra-massive or a stable NS, remain possible. 

X-ray observations have a prime role in constraining the merger final product, as newly born NS can be bright sources of X-ray radiation \citep{verbunt96,kargaltsev13,metzger14}.
Such radiation is initially blocked by the merger ejecta surrounding the remnant \citep{metzger14} but, as the ejecta expand and cool down, observations can peer down at the central compact object. 
 Past works \citep{Troja2018, Lazzati2018, Margutti18} already showed that X-ray emission from GW170817 is well described by standard afterglow synchrotron radiation, produced by the interaction of a relativistic outflow with a low-density ($n \lesssim 0.001$\,cm$^{-2}$) ambient medium at large radii ($\approx$10$^{18}$\,cm) from the central power source.  
Any contribution from the central compact source must  therefore be comparable to the GRB afterglow luminosity or higher  in order to be detected.

In this paper we report the multi-wavelength afterglow data taken with ATCA, HST, XMM-Newton and \textit{Chandra} around the  broad local maximum in X-ray brightness reached at day $\approx$ 150.  
We discuss the model of a structured relativistic jet \citep{AloyJankaMueller2005,Lazzati2017,kathi18,Xie2018} launched by the merger remnant and seen at a large viewing angle from its axis. The underlying engine is a long-lived magnetized NS, which injects energy into the relativistic outflow and the sub-relativistic ejecta \citep{metzger18,yu17,ai18}
We  discuss the consistency of such scenario with the broadband data, 
including the kilonova properties, the afterglow long-term evolution, and 
the possible presence of short-term variability in the X-ray data. 

While previous comparisons assumed a  magnetic dipole
spindown loss \citep{pooley18}, we consider the GW-dominated spindown regime,  that is expected from a NS newly born from a merger.    Constraints on the NS configuration, with particular regard to its  magnetic field and ellipticity  are derived. Implications on the NS mass, EoS and future observing strategy, with particular regard to X-ray observations, are briefly discussed.

\section{Observations}
 \subsection{X-rays}\label{sec:Xrays}

A log of X-ray observations around the broad local maximum in X-ray brightness reached at day $\approx$ 150  is reported in Table~\ref{tab:xray}.  Earlier observations were reported in  \citet{Troja2017,Troja2018,Davanzo2018,haggard2017}, while the most recent in \citet{Troja18b}. {\it Chandra} data were reduced in a standard fashion using the CIAO v4.9 and the latest calibration files. Source counts were extracted from a circular region containing 92\% of the encircled energy fraction, whereas the background contribution was estimated from nearby source-free regions.
We verified that none of the observations was affected by high levels of particle background. 

{\it XMM-Newton} data were processed using SAS v16.1.0 and the most recent calibration files. Periods of high background were excluded from the analysis. The native astrometry was refined by matching the positions of 5 bright X-ray sources with their optical counterparts in the GSC v2.3.2 catalogue\citep{Lasker08}. In order to minimize the contribution from contaminating X-ray sources, a small aperture of 5'' was used to extract the source counts. 

X-ray spectra were binned in order to have at least one count per energy channel and fit within the XSPEC v12.8.2 package by minimizing the C-statistics \citep{Cash79}. To convert the observed count-rates to flux values we adopted a spectral index $\beta =0.58$ as derived from the broadband spectral energy distribution \citep{Troja2018,Troja18b}. 

\subsubsection{Temporal analysis}

 As shown in Table~\ref{tab:xray}, the X-ray observations performed around 160 d post-merger were split into several exposures spread over a period of a week. This allowed us to search for variability on short time-scales. 
During the first two {\it Chandra} observations, performed at 153 and 157 d, we measure a total of 89 source counts in 48 ks of exposure. 
In the last two observations, performed at 161 and 165 d, 
the count rate is lower, and we measure a total of 37 source counts in 37 ks of exposure. For a constant source, the Poissonian probability for such fluctuation is  $\approx$3.3 $\sigma$.

\begin{figure}
 \includegraphics[width=0.95\columnwidth]{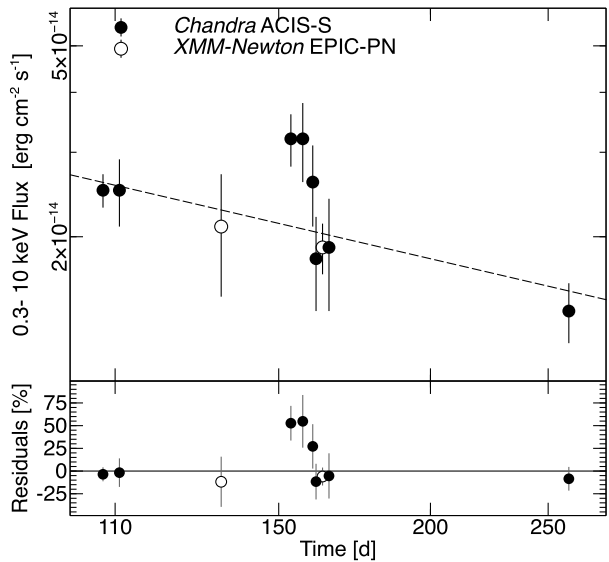}
\caption{ X-ray afterglow of GW170817 after 100 days (top panel). Vertical error bars are 1 $\sigma$.
The dashed line shows the power-law fit model.  Fractional residuals are shown in the bottom panel. }
    \label{fig:xray}
\end{figure}

\begin{figure*}
 \includegraphics[width=1.9\columnwidth]{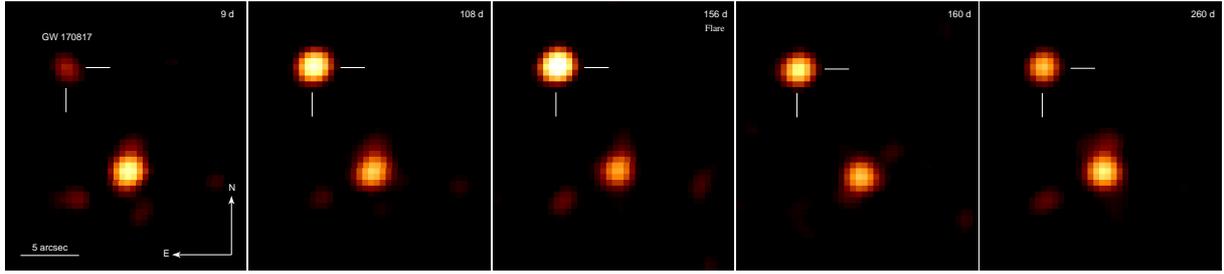}
\caption{ X-ray afterglow of GW170817.
Images are background subtracted, corrected for exposure, and smoothed with a Gaussian function with $\sigma$=1.5\arcsec. The X-ray emission from GW170817 is seen to slowly evolve with time. However, a rapid decrease in brightness is observed between 156d and 160d after the NS merger.
During this interval, the X-ray count rate decreases by a factor of 1.7. 
Between 160~d and 260~d, it decreases by a factor of 1.3.}
    \label{fig:ximage}
\end{figure*}

By including the adjacent X-ray data (Table~\ref{tab:xray}) we obtain a similar significance of the temporal feature,   an X-ray flare of a few days duration, peaking at $\approx$ 155 d.   In order to estimate this value,  we fit the X-ray data with a simple power-law model (Figure~\ref{fig:xray}) and used this best fit continuum as input for a set of 10,000 Monte Carlo simulations.  For each simulated dataset, we searched for statistical fluctuations mimicking a flare, derived the likelihood value of the two models (continuum vs continuum+gaussian flare) and calculated their ratio. 
Only in 11 cases we found a ratio lower than the observed value. We therefore conclude that the probability of a statistical fluctuation resembling a flare-like feature as significant as the one observed at 160 d is $\approx$10$^{-3}$.
 Images showing the evolution of the afterglow are presented in Figure~\ref{fig:ximage}.

\subsection{Optical observations}\label{sec:optical}

We obtained two late-time epochs of imaging (PI: Troja) with the Hubble Space Telescope.  Images were taken with the UVIS detectors of the Wide-Field Camera 3 (WFC3). Data were reduced in a standard fashion using the Hubble Space Telescope CalWF3 standard pipeline \citep{hst01}, and the astrodrizzle processing \citep{hst02}.  The final pixel scale was  0.3$^{\prime \prime}$. 

\begin{figure}
\centering
 \includegraphics[width=0.95\columnwidth]{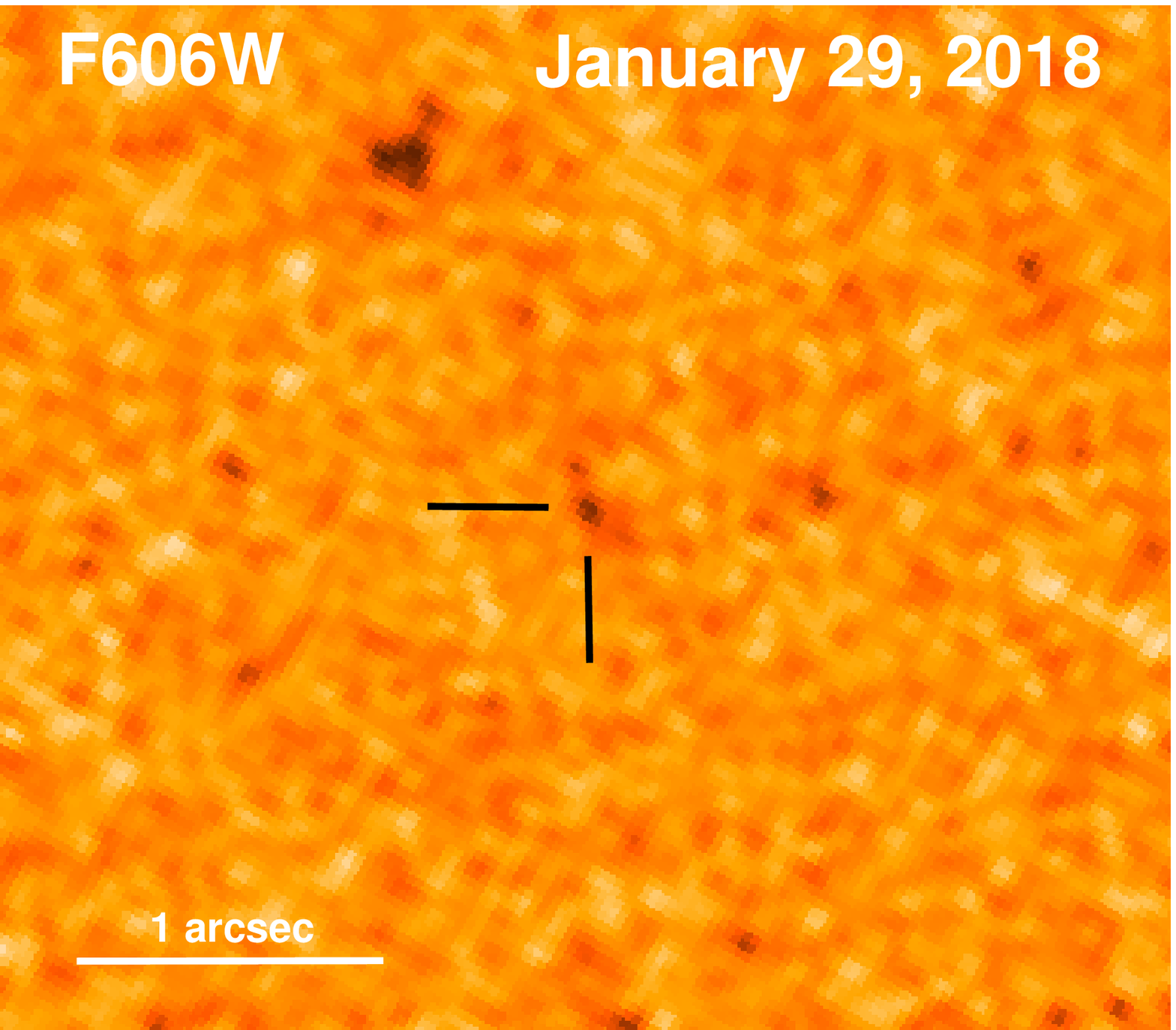}
  \includegraphics[width=0.95\columnwidth]{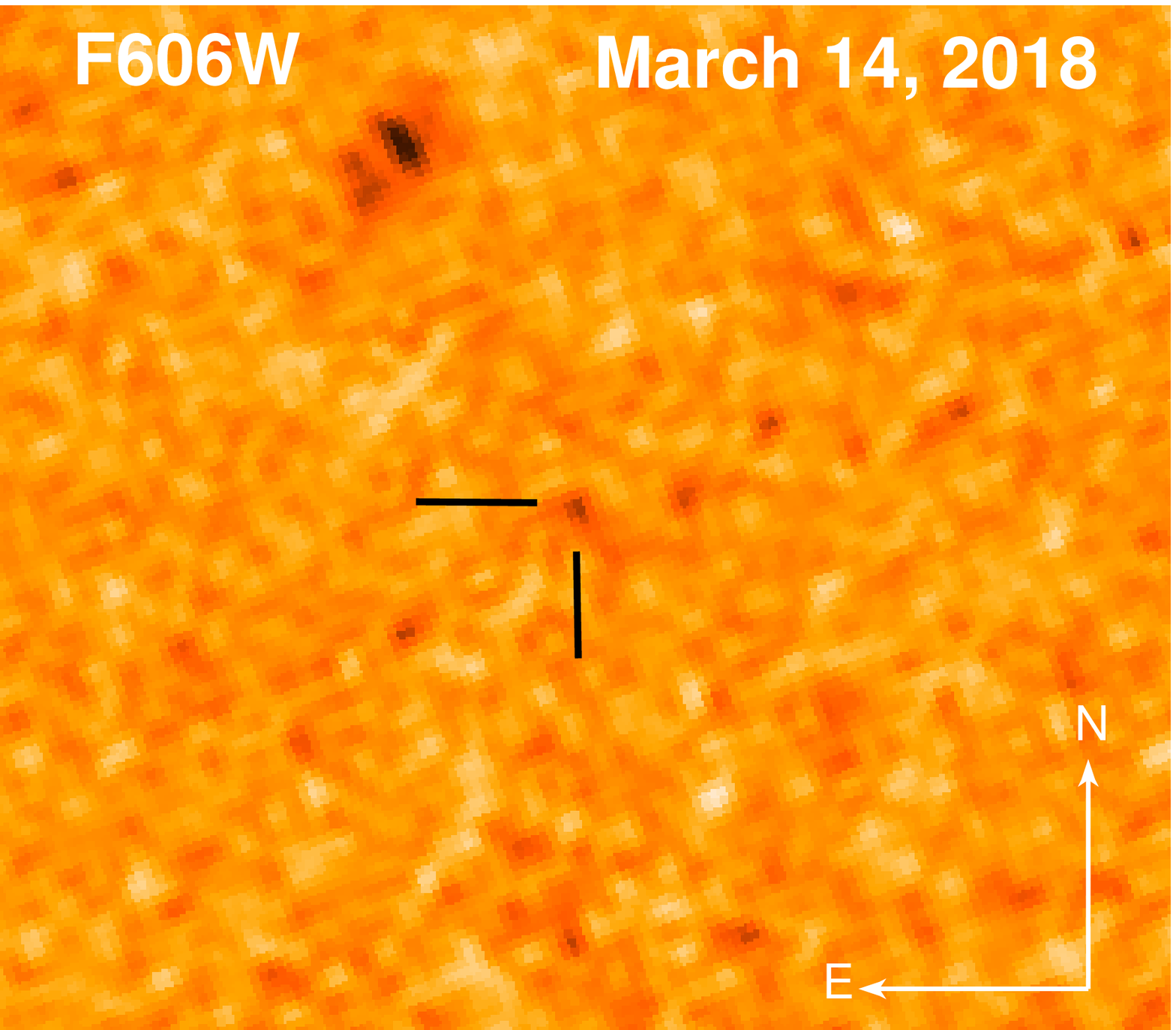}
\caption{ Optical afterglow of GW170817 at 166 d (top panel)
and 209 d (bottom panel) after the merger. 
Images are galaxy subtracted and smoothed with a Gaussian function of a 2 pixels width. 
The source position is indicated by the lines. }
    \label{fig:hst}
\end{figure}

To subtract the galaxy light we used a median filter with window size
of 15 times the FWHM of PSF of stars (3.3 pixels), large enough to remove the structure of the galaxy but not point sources like the afterglow.
The residual images are shown in Figure \ref{fig:hst}.
The GRB afterglow is weakly detected during our first epoch (top panel),  whereas in our later epoch the source, although marginally visible in the residual image (bottom panel),  is of low (< 2 $\sigma$) significance. 
Images were analyzed using PSF-photometry based on DAOPHOT tasks under IRAF. 
 We estimated an observed magnitude $F606W$=26.7 $\pm$0.4 AB mag in our first epoch, and $F606W$> 26.6 AB mag in our last epoch. 
Our final photometry is listed in Table~\ref{tab:hst}. 
Earlier observations were reported in \citet{Lyman18} and  \citet{Margutti18}.

The phenomenological model of \citet{Dobie18} predicts a continued rise of the radio afterglow up to 150 d. If this model were extended to 
the optical wavelengths, it would be inconsistent with our optical data,  that instead favor a smoother, flatter turn-over of the optical light curve.

\subsection{Radio observations}\label{sec:radio}

The target source was observed with the Australia Telescope Compact Array 
(ATCA) at five different epochs under programs CX394 (PI: Troja) and CX391 (PI: Murphy).
In order to bootstrap the flux density scale the standard source 1934-638 was observed in all epochs. The phase calibrators 1245-197 (first two epochs) and 1244-255 (last three epochs) were used to compute the complex gains.
All the data sets were flagged, calibrated and imaged using standard 
procedures in the data reduction package MIRIAD. 
In order to maximize the results the 5.5 and 9 GHz data were imaged using a robustness parameter value of r=0.5 (1st and 2nd epochs) and r=-0.5 (4th and 5th epochs). 
Flux measurements for all epochs are reported in Table~\ref{tab:radio}. 
Whereas our measurements at 9 GHz are generally consistent with \citet{Dobie18}, 
the derived fluxes at 5.5 GHz are systematically lower, and in better agreement with the 
VLA measurements at similar epochs \citep{Margutti18}. 
Additional observations were reported in \citet{Mooley2018,Troja2018,Margutti18} and \citet{Troja18b}.

\begin{table}
 	\caption{ HST observations of GW170817. Upper limits are 3\,$\sigma$. Magnitudes are corrected for Galactic extinction using \mbox{E(B--V)}=0.105 \citep{sf11}.}
 	\label{tab:hst}
 	\begin{tabular}{ccccc}
    \hline
    $T-T_0$ & Instrument & Filter & Exposure & AB mag \\
     (d) &    &  & (s) & \\
        \hline
           166 &  WFC3/UVIS & F606W & 2372 &  26.4 $\pm$0.4  \\
           209 &  WFC3/UVIS & F606W & 2432 & $<$26.3  \\
       \hline
    \end{tabular}
\end{table}

\begin{table*}
 	\caption{ ATCA observations of GW170817. Errors are 1~$\sigma$. Upper limits are 3\,$\sigma$}
 	\label{tab:radio}
 	\begin{tabular}{cccccc}
    	\hline
    $T-T_0$ & Frequency & Bandwidth & Configuration & Exposure & Flux  \\
     (d) &    (GHz)       &  (GHz)      &            & (hrs) &  ($\mu$Jy) \\
     \hline    
       125   &  5.5 &  2.0 & 6C & 10.5  &  72\,$\pm$\,9 \\
             &  9.0 &  2.0 & 6C &   "       &  72\,$\pm$\,9  \\
       149   &  5.5 &  2.0 & 6C & 10.5  &  79\,$\pm$\,8   \\
             &  9.0 &  2.0 & 6C &  " &  50\,$\pm$\,7 \\
       160   &  19  &  4.0 & 750A & 10.5  &   $<36$ \\
       168   &  5.5 &  2.0 & 750A & 6.5   &   $<87$  \\
             &  9.0 &  2.0 & 750A & "   &   $<126$  \\
       182   &  5.5 &  2.0 & 750B & 9.5   &     81\,$\pm$\,16 \\
             &  9.0 &  2.0 & 750B & "     &     54\,$\pm$\,11 \\
       221   &  5.5 &  2.0 & EW352 & 12.0 &     60\,$\pm$\,12 \\
             &  9.0 &  2.0 & EW352 & 12.0 &    $<30$          \\   
        \hline
    \end{tabular}
\end{table*}

\section{A long-lived magnetized NS as the merger remnant}

GW observations constrain the mass of the remnant to $<2.8~M_{\odot}$, but do not break the degeneracy between a NS and a BH \citep{LIGO-product}. 
Depending on the unknown NS equation of state and the spindown history, a supra-massive (up to 20\% more massive than the maximum mass of a non-spinning neutron star \citep{breu16}) or even a permanently stable NS can survive after the merger. Here we discuss the implications of the observations for such a model.

\subsection{Consistency with broad band observations}
In order to accommodate the available electromagnetic observations, the merger product should have a weak poloidal magnetic field \citep{SwiftGW,ai18}. During the spin-down process (either due to magnetic dipolar radiation or secular GW radiation), a continuous Poynting-flux-dominated outflow is launched and adds energy into the ejecta. 
The dipolar poloidal magnetic field at the NS surface should be below $\approx 10^{12}$ G in order to satisfy the upper limits set by the broadband observations, including the prompt $\gamma$-rays, the kilonova emission and the long-term X-ray, optical, and radio afterglow \citep{ai18,pooley18}.  
Indeed, the latest claim of a BH merger product \citep{pooley18} suggests that the electromagnetic luminosity from the spin-down energy of a rapidly spinning NS ($2\times 10^{52}$ erg) is ruled out by the data. On the other hand, a newly-formed NS likely possesses a large ellipticity so that secular gravitational wave loss is expected to remove a significant amount of its initial spin energy \citep[e.g.][]{dallosso07,dallosso09,fan13,gao16}. The argument of \cite{pooley18} is removed when gravitational wave spindown is properly taken into account.

The X-ray luminosity of a spinning magnetized NS is  given by the 
energy input into the surrounding medium from electromagnetic losses \citep{Lasky16}.  
\begin{equation}
	L(t)=\frac{\eta B_p^2 R^6\Omega(t)^4}{6c^3}
 \label{lem}
\end{equation}
where $\Omega(t)$ is the solution of the spin down equation (eq.1. of \citep{Lasky16}, $B_{p}$  the dipole component of 
the magnetic field,  $R$  the neutron star radius, respectively. The efficiency $\eta\le1$ accounts for  converting spin-down energy into electromagnetic radiation, through the X-ray channel. 
 $\Omega(t)$ reflects the dominant spin down losses, either  emission of GW or dipole radiation, that are characterized by the  time scales:
 \begin{equation}
 \tau_{gw}=\frac{5c^5}{128GI\epsilon^2\Omega_0^4}= 9 \times 10^5 \epsilon_{-4}^{-2} I_{45}^{-1} P_{-3}^4\ {\rm s}
 \label{tauGW} , 
 \end{equation}
\begin{equation}
 \tau_{em}=\frac{3c^3I}{B_p^2R^6\Omega_0^2}= 2\times 10^9I_{45}R_6^{-6}B_{p,12}^{-2}P_{-3}^2\ {\rm s},
 \end{equation}
where $\epsilon$ is the ellipticity and $P$ the period.
When $\tau_{gw}<1/2 \tau_{em}$, gravitational wave emission dominates spin down until a time  
\begin{equation}
\tau_*=\frac{\tau_{em}}{\tau_{gw}}\left(\tau_{em}-2\tau_{gw}\right).\label{taustar} 
\end{equation}

For $t<\tau_*$ the X-ray luminosity follows
\begin{equation}
L(t)=L_0 \left(1+\frac{t}{\tau_{gw}}\right)^{-1}
\label{Lgw}
\end{equation}
where
\begin{equation}
 L_0=\frac{\eta I \Omega_0^2}{2\tau_{em}}=10^{40} \eta_{-3} R_6^6 B_{p,12}^2 P_{-3}^4 \space   {\rm erg\ s^{-1}}
 \label{Lem0}
 \end{equation}

Comparison with present observations by \citet{pooley18} assumed that   electromagnetic radiation dominates spin-down. In such a case the luminosity follows
\begin{equation}
L(t)=L_0 \left(1+\frac{t}{\tau_{em}}\right)^{-2}.
\label{Lem}
\end{equation}

However, in the GW-loss dominated regime, the luminosity becomes  a factor
$\propto t/\tau_*$ lower,  thus relaxing the constraints derived from  observations.
This condition applies when $\frac{\tau_{gw}}{\tau_{em}}<<1$ that is satisfied when
\begin{equation}
\epsilon_{-4}>2\times 10^{-2} I_{45}^{-1}R_6^{-3}P_{-3} B_{p,12}
\end{equation}
and  the corresponding X-ray flux from equation~\ref{Lgw} (assuming D=40 Mpc) is given by
\begin{eqnarray}
F_X= \left\{
\begin{array}{ll}
5\times 10^{-14} \eta_{-3} R_6^6 B_{p,12}^2 P_{-3}^{-4} \hspace{35pt}  t<\tau_{gw}   \\
4\times 10^{-15} \eta_{-3} R_6^6 B_{p,12}^2 I_{45}^{-1} \epsilon_{-4}^{-2} t_7^{-1} \hspace{10pt}   t>\tau_{gw}  
\end{array}
\right.
\label{fx}
\end{eqnarray}
with the flux  in ${\rm erg\ cm^{-2}\ s^{-1}}$.

We require this flux to be consistent with X-ray observations at t>100 days, when the ejecta are optically thin.  This sets a first condition on $\tau_{gw}<100 \rm d$, i.e. $\epsilon_{-4}>3\times 10^{-1} I_{45}^{-1/2}P_{-3}^{2}$. A second condition follows by requiring that the flux at $t>100$ d  be lower than the observed one:
\begin{equation}
\epsilon_{-4}>0.3 R_6^{3}I_{45}^{-1/2}\eta_{-3}^{1/2} B_{p,12}=0.5 \eta_{-3}^{1/2} B_{p,12}.
\end{equation}
This equation   provides the tighter constraint on $\epsilon$ for the assumed parameter of the NS ($M=2.1M_{\odot}$, $R_6=1.2$ and $I_{45}=2$), and allow us to conclude that a NS with ellipticity $\epsilon \gtrsim 10^{-5}$ and  a poloidal field $B_{12}\gtrsim 0.1$ is fully consistent with the X-ray dataset collected so far.

The required ellipticity can be produced by a strong toroidal component of the magnetic field that develops from the differential rotation expected from a NS born from the merger \citep{rezzolla18,giacomazzo15}. The strong magnetic field gradient is expected to deform the star with an ellipticity that can be approximated by $\epsilon\approx 10^{-5}\left(\frac{B_t}{3\ 10^{15}G}\right)^2 $ \citep{cutler02}, where $B_t$ is the toroidal component of the field.
Another  viable mode for developing ellipticity involve the so called bar mode instability \citep{Corsi09}, that can  produce $\epsilon$ as large as $10^{-3}$ \citep{Lasky16}.

\subsection{Alleviating  the requirements on kilonova ejecta}
A long-lived NS is not only allowed, but is also helpful to interpret some of the data. 
Energy injection to the kilonova from such a remnant indeed helps to interpret the kilonova properties without invoking extreme parameters \citep{yu17,li18}. 
The remnant NS deposits extra energy to power the kilonova emission \citep{yu13,kasen15,murase18}. This helps to account for the early peak and high luminosity of the ``blue kilonova'' \citep{SwiftGW}, otherwise difficult to explain with standard model parameters \citep{Troja2017,li18}. 
Indeed, a NS with initial spin-down luminosity of $\sim 3.4 \times 10^{44} \ {\rm erg \ s^{-1}}$ at $500$~s and a luminosity evolution $\propto t^{-1}$ (gravitational wave spindown dominated regime) can account for the multi-wavelength evolution of AT2017gfo without the need of introducing a large amount of ejecta mass and an unreasonably small opacity \citep{li18}. With these parameters, the spin-down luminosity at $\sim 1$ day is $\sim 2 \times 10^{42} \ {\rm erg \ s^{-1}}$, too low to significantly affect the opacity of the merger ejecta \citep{metzger14}.
This satisfies the observational constraint of a ``red kilonova'' component as well as the spectral features of lanthanides elements \citep{kasen17,Pian17, Troja2017}.

\subsection{A NS as the central engine of short GRBs}
Previous criticisms to  a long-lived NS remnant included the apparent difficulty of producing a short GRB in a neutron star engine \citep{metzger18,margalit17}.
Mechanisms to produce a short GRB in a neutron star central engine without the introduction of a black hole have been discussed in the literature, including early accretion \citep{metzger08} or magnetic activities due to differential rotation \citep{fan13}. A good fraction of short GRBs are found to possess an extended ``internal plateau''\citep{Troja07}, which suggested the existence of a supra-massive or stable neutron star \citep{rowlinson13,lv15}. Interpreting these features within the neutron star engine model indeed require significant energy loss in the gravitational wave channel \citep{gao16}.

\begin{figure}
\includegraphics[width=\columnwidth]{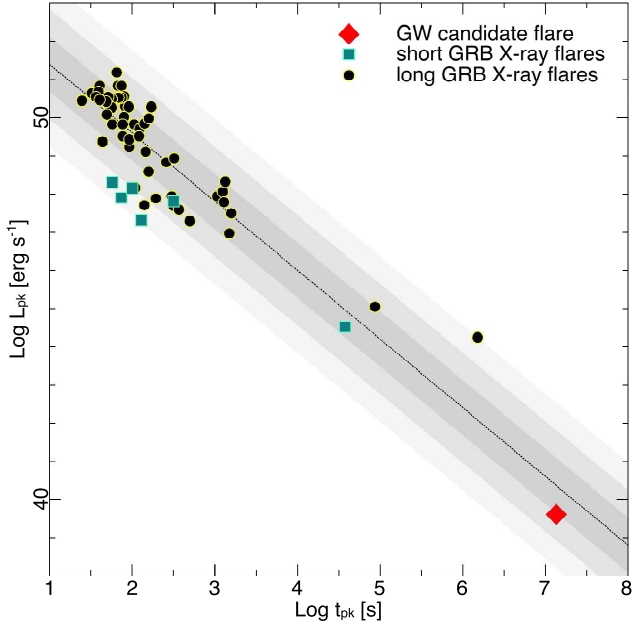}
\caption{Comparison with X-ray flares in GRB afterglows. The luminosity and peak time of the candidate X-ray flare in GW170817 (red diamond) follow the trend observed in GRB X-ray flares. The best-fit relation for GRB X-ray flares from \citealt{Bernardini2011} 
is shown by the dashed line. The shaded areas shows the 1$\sigma$ (dark grey), 2$\sigma$ and 3$\sigma$ (light grey) regions.}
\label{fig:xrayflares}
\end{figure}


\begin{figure}
 \includegraphics[width=0.9\columnwidth]{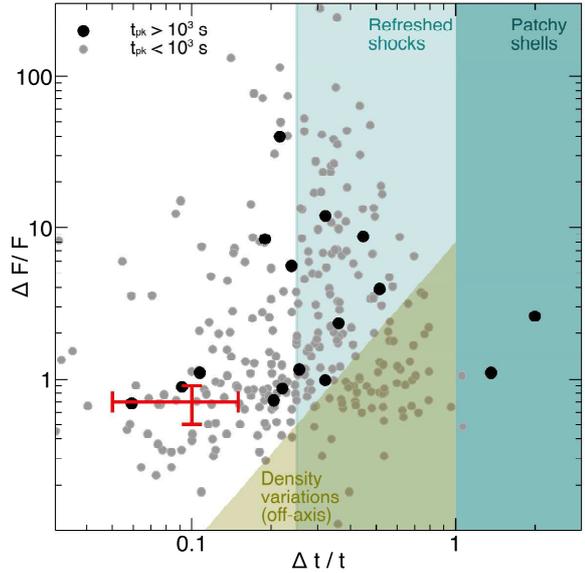}
\caption{Ioka diagram for X-ray flares.
X-ray flares in GRBs (circles) and GW170817 (red data point) are shown. The horizontal error bar reports the uncertainty in the flare duration due to the sparse sampling. The shaded areas show the regions allowed by afterglow models \citealt{ioka05}.
More detailed shock models exclude density variations below $\Delta t/t \lesssim 1$. Most X-ray flares, including the one observed in GW170817, lie outside these regions.}
    \label{fig:ioka}
\end{figure}

\subsection{Late time X-ray variability and a long-lived magnetized NS}

On top of the overall trend produced by the relativistic outflow, X-ray monitoring of the source  exhibited a candidate X-ray flare. Between January 17 and January 28 2018, six consecutive X-ray observations  displayed a variation by a factor $\approx 1.7\pm0.2$ in the X-ray flux (Figure~\ref{fig:xray}). 
 The sparse sampling of X-ray observations  prevents a search of  similar temporal variations at other epochs.

X-ray flares are erratic temporal features, commonly seen in GRB afterglows, and often attributed to a re-activation of the central power source \citep{Burrows05,zhang06,Chincarini07}. Their emission peaks in the X-ray range, and is often undetected at other energies \citep{Troja15}.
The X-ray observations of GW170817 do not sample the entire temporal profile of the candidate flare, thus preventing a detailed comparison with the population of GRB X-ray flares. Nevertheless, some of its basic properties can be estimated. 
The similar fluxes measured at 155 and 157 days, followed by a rapid decay phase, suggest that the emission peaked around those dates. 
The peak time, $t_{pk} \approx$ 156 d, and peak luminosity, $L_{pk} \approx 2 \times 10^{39}$\,erg\,s$^{-1}$, fall within the expected range of values derived by extrapolating the distribution of GRB X-ray flares \citep{Bernardini2011} to later times (Figure~\ref{fig:xrayflares}).
We conservatively estimate the flare width as the time interval between the two
X-ray observations consistent with the baseline continuum, that is t$_1$=137 d and $t_2$=161 d, which yield $\Delta t \lesssim$24~d and $\Delta t / t \lesssim$0.15. The decay phase observed after 157 d places a lower limit of $\Delta t \gtrsim$6~d and $\Delta t / t \gtrsim$0.04.
Such rapid variability places our candidate flare in a region that is excluded by most afterglow models (\citet{ioka05}, Figure~\ref{fig:ioka} and  Appendix). 

Most naturally, and in analogy with X-ray flares in GRBs,  the  candidate flare observed in GW170817 is likely related to a central engine that is still active at late times. 
This scenario receives support from the so-called ``curvature effect'' test \citep{liang06}. Any flare is bound  to follow a temporal decay shallower  than $\alpha = 2 + \beta$ \citep{kumar00}, where $F_{\nu}\propto t^{-\alpha} \nu^{-\beta}$ and, in our case, $\beta\sim0.58$ \citep{Troja2018}. 
By using the merger time as our reference time $T_0$, the measured power law decay slope of the flare is $\alpha \sim 9.9$, greater than the predicted value. This is likely due to a mis-identified zero time $T_0$ \citep{zhang06}. By imposing that $\alpha = 2 + \beta =$2.58 and fitting for $T_0$, we find that  $T_0$ is $116^{+11}_{-26}$ d. This marks the beginning of the flare, which is consistent with our hypothesis that the central engine was reactivated to power the flare.

If the final merger product is a BH, then its re-activation could be due to either fallback accretion \citep{Rosswog2007} or disc fragmentation \citep{perna06}. In the former scenario, the total fallback power declines as $t^{-5/3}$ 
and, for typical ejecta masses of NS mergers, is $\lesssim$10$^{39}$ erg\,s$^{-1}$ at 160 d after the merger. This is comparable to the observed X-ray luminosity, and would therefore require an unrealistic radiative efficiency in order to accommodate our observations. 
The latter scenario needs the accretion disc to survive for months, which is not expected based on our understanding of NS mergers \citep{perna06}.

As discussed above, a  supra-massive  or even a permanently stable NS can survive after the merger. Due to its initial rapid differential rotation, this post-merger NS likely has a strong toroidal component of the magnetic field and possibly also a strong poloidal component \citep{thompson93}. 
The untwisting of the toroidal magnetic field may give rise to an abrupt injection of outflows with enhanced wind luminosity {with a mechanism similar to GRB X-ray flares \citep{dai06} or bursts and flares of soft gamma-ray repeaters \citep{thompson01}. The internal magnetic dissipation of such an outflow \citep{zhangyan11} would give rise to flaring emission observable in X-rays. 
We estimate the toroidal component of the magnetic field as follows. The total isotropic-equivalent energy of the flare is in the range $7 \times 10^{44} \ {\rm erg} < E_{\rm flare} < 3 \times 10^{45} \ {\rm erg}$. This is much smaller than the total spin energy of a new-born millisecond pulsar. If one exclusively attributes the flare energy to the NS magnetic field energy, then $B^2 R^3/6 \gtrsim 3 \times 10^{45}$ erg. Therefore, the required toroidal magnetic field stored in the NS must be $B_{\rm t} \gtrsim 10^{14}$ G, which is reasonably expected \citep{thompson93}.

The dipolar poloidal magnetic field at the NS surface should be  $\approx 10^{12}$ G in order to satisfy the upper limits set by the broadband observations.  
Such a high-toroidal-$B$ and low-poloidal-$B$ NS is analogous to the source SGR 0418+5729 \citep{tiengo13} that emits magnetar flares  but has a dipolar magnetic field \citep{rea10} lower than $7.5\times 10^{12}$ G. 

\subsection{Internal magnetic dissipation in the NS outflow}

Since the $\alpha=2+\beta$ ``curvature effect'' test \citep{kumar00,liang06} suggests restarting of the central engine at the flare, the X-ray emission likely originates from a radius $R_{\rm flare} \sim \Gamma_{\rm flare}^2 c \Delta t_{\rm decay} \sim (2.6\times 10^{18} \ {\rm cm}) (\Gamma_{\rm flare}/10)^2 (\Delta t_{\rm decay}/10 \, {\rm d})$, where $\Delta t_{\rm decay} \sim 10$ d is the decay time scale of the flare. At $\sim 150$ d after the merger, the external shock blastwave has moved to a distance $R_{\rm blast} \sim \Gamma_{\rm blast}^2 c t \sim (6.2\times 10^{18} \ {\rm cm}) (\Gamma_{\rm blast}/2)^2 (t/150 \, {\rm d})$ from the central engine.   Around the flare, $\Gamma_{\rm blast} \approx 1 / \theta_{v}\approx 2$  since the flare happens
around the light curve turnover point when the jet tip is visible . Therefore the flare emission is ``internal'' if the Lorentz factor of the emitting material is $\approx 10$. This is consistent with various constraints that GRB X-ray flares have a lower Lorentz factor than GRB themselves \citep{yi15}. The trigger of the flare may be through collision-induced magnetic reconnection and turbulence \citep{zhangyan11,deng15} or an external-pressure triggered kink instability \citep{lazarian18}. Either way, an enhanced release of the Poynting flux energy due to reconnection is induced, giving rise to the flare emission.

 According to the above estimate, the flare emitting region is outside the radius of the non-relativistic merger ejecta, $R_{ej} \lesssim (1.2\times 10^{17} {\rm cm}) (\beta/0.3)(t/150 \rm d)$. This can be understood as follows: in the observer's viewing direction, there is already a funnel  opened by the earlier relativistic ejecta that powered the prompt  and afterglow emission of GRB 170817A. With continuous energy injection from a spinning-down NS, the funnel would 
remain open 
 so that the newly ejected enhanced Poynting flux can penetrate through the  non-relativistic merger ejecta and reach the large radius where X-ray emission is released. 

In order to see whether the funnel remains open, one can compare the pressure of the non-relativistic merger ejecta and the  comoving-frame magnetic pressure of the long-lasting pulsar wind. Suppose that the central engine spindown luminosity evolves with time as 
\begin{equation}
L(t) \propto t^{-q},
\label{q}
\end{equation}
the comoving-frame magnetic field strength of the pulsar wind may be estimated as $B' \propto L^{1/2} R^{-1} \Gamma^{-1}$, so that the magnetic pressure scales as $p_B = B^2/8\pi \propto t^{-q} R^{-2}$.  Here we have assumed that the Lorentz factor of the pulsar wind, $\Gamma$, does not evolve significantly with time. The gas pressure of the ejecta, on the other hand, scales as $p \propto \rho^{5/3} \propto R^{-10/3} \propto t^{-10/3}$ assuming adiabatic evolution and no radial spreading of the ejecta. Radiative loss and radial spreading would further steepen the decay. We consider the competition between $p_B$ and $p$ at the radius of the ejecta, so that $R \propto t$. One can then compare $p_B \propto t^{-(2+q)}$ and $p \propto t^{-10/3}$. For a low-$B$ pulsar, the spindown time scale is long. One may make a connection between the spindown time scale and the turn-over time of X-ray emission ($\sim 160$ d). Before this time, one has $q$ either 0 (dipole-spindown-dominated) or 1 (secular-GW-spindown-dominated). For both cases (and any intermediate value of $q$), the decay slope of $p_B$ is shallower than the decay slope of $p$. This suggests that the funnel would remain open, and likely would widen as a function of time.

\begin{figure}
 \includegraphics[width=1.0\columnwidth]{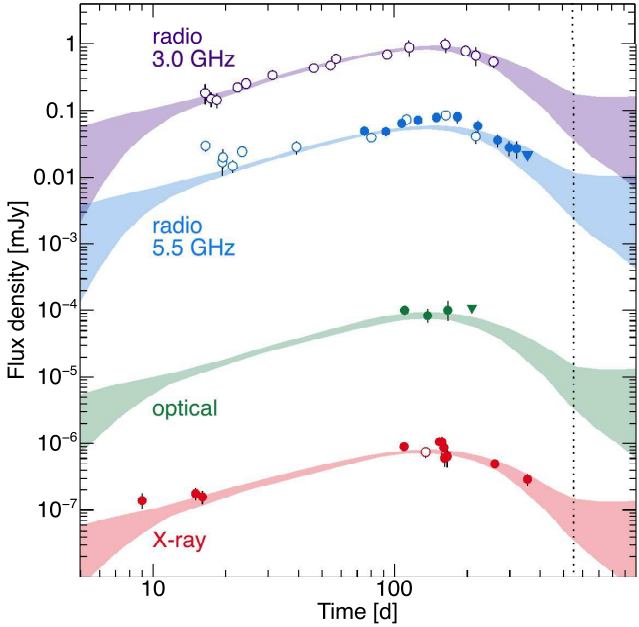}
\caption{Afterglow temporal evolution for GW170817. 
The multi-wavelength dataset is compared with a Gaussian jet model with the addition of energy injection from the pulsar  as described in the text.   The width of each model curve indicates the 68\% range of confidence. The radio data and model at 3 GHz are scaled by a factor of 10. Energy injection from the pulsar has a negligible effect on the observed afterglow, and may cause a flattening only at late times  ($\gtrsim 2$ yr, vertical dotted line).}
 \label{fig:lc_ei}
\end{figure}

\begin{figure}
 \includegraphics[width=\columnwidth]{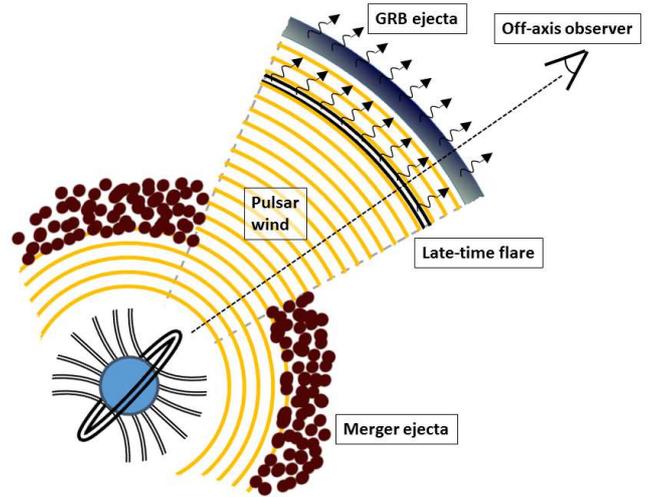}
\caption{Scheme of the model. A structured jet is launched by rapidly spinning long-lived NS with a strong differential magnetic field. The structured jet fully accounts for the broad band non-thermal continuum. 
The pulsar wind would  provide additional energy injection to the non-relativistic merger ejecta that produce the kilonova features.
Abrupt magnetic reconnection of the strong toroidal component  launches brief relativistic outflow that produce X-ray flares via internal collision-induced magnetic dissipation. The magnetic axis is perpendicular to the spin axis, which is likely
the outcome of the spin-flip instability for a magnetically-distorted
neutron star \citep{jones76,cutler02,Lasky16,dallosso18}. The GW spindown is significant in
such a configuration, as assumed in our analysis. }
    \label{fig:cartoon}
\end{figure}

\subsection {Effects of energy injection on the afterglow}
\label{injection_section}

Starting on August 26 2017 \citep{Troja2017}, X-ray light from the transient GW170817 is being detected by NASA's Chandra X-ray Observatory and, more recently, by ESA's XMM-Newton satellite \citep{Davanzo2018}. This X-ray emission brightened by a factor of five  during the first three months following the NS merger \citep[e.g.][]{Troja2018,Margutti18}, reaching a luminosity at peak of $\approx$4$\times$10$^{39}$\,erg\,s$^{-1}$. The temporal evolution of the X-ray signal can be described by a power-law rise, $L_X$$\propto t^{0.9}$, followed by a smooth turn-over $\approx$100 days after the NS merger and then a phase of rapid decay (\citealt{Troja18b}).
A similar behavior is displayed by the radio and the sparser late-time optical data  \citep{Mooley2018,Dobie18,Lyman18} and is well described by models of structured jets \citep{Troja18b,Lazzati2017,Margutti18,Xie2018,Lyman18}. 

The existence of a central engine pulsar could  provide additional energy injection to the afterglow blastwave, potentially altering the evolution of the forward shock and the ensuing electromagnetic emission.
Energy injection into a blastwave by an underlying pulsar has been extensively studied \citep{dai98,zhang01}. For an engine satisfying Eq.(\ref{q}), in the spectral regime below $\nu_c$ (where the X-rays seem to lie in), the forward shock flux scales as \citep{zhang01,zhang06}
\begin{equation}
F_\nu \propto t^{(1-q)-\frac{(p-1)(2+q)}{4}},
\end{equation}
which is valid for $q \leq 1$. The broad-band afterglow spectral index of GW170817 is $p\sim2.17$. The  observed $F_\nu \propto t^{0.9}$ rise of the afterglow demands $q \sim -0.4$, which is out the scope of the pulsar model.  For $q = 1$ (relevant for secular-GW-spindown-dominated case), energy injection is essentially negligible. This suggests that energy injection from the NS can at most partially contribute to the observed afterglow emission, and additional energy injection, either from high latitudes of a structured jet or from a stratified ejecta outflow, is needed to reproduce the rising phase of the GW170817 afterglow. 

We verified that for $q=0$ (relevant for dipolar-spindown-dominated phase), the engine injection from the pulsar does not alter the afterglow emission provided $L_0 < 4\times 10^{44}$erg/s. We expanded the Gaussian jet model to include isotropic energy injection of the form $L(t) = L_0 (t/t_0)^{-q}$ until a stop time $t_s$.  
To fit this model to the data we perform Bayesian parameter estimation by sampling the posterior probability distribution with a Markov-Chain Monte Carlo (MCMC) package\citep{Foreman-Mackey2013}.  

When included in an MCMC run, we find that the energy injection must be a sub-dominant component and obtain an upper limit $L_0 < 4\times 10^{44}$ erg~s$^{-1}$ with 95\% confidence. The $q$ and $t_s$ parameters are unconstrained, and the other parameters of the jet 
as presented in \citet{Troja18b} 
are unchanged.
While energy injection from the pulsar has a negligible effect on the observed afterglow, it may cause a flattening  at late times  ($\gtrsim 2$yr, Figure~\ref{fig:lc_ei})}.

\section{Conclusions}

GW observations constrain the mass of the remnant to $<2.8~M_{\odot}$, but do not break the degeneracy between a NS and a BH \citep{LIGO-product}.  
 Depending on the unknown NS equation of state and the spindown history, a supra-massive (up to 20\% more massive than the maximum mass of a non-spinning neutron star \citep{breu16}) or even a permanently stable NS can survive after the merger.
 Due to its initial rapid differential rotation, this post-merger NS likely has a strong toroidal component of the magnetic field and possibly also a strong poloidal component \citep{thompson93}. 
 Previous criticism to a NS remnant was based on the high X-ray luminosity expected from a spinning NS, found to be marginally consistent with observations only for a relatively small value of the dipole magnetic field \citep{pooley18}. However, the aforementioned argument was based on the assumption that the spin down losses are dominated by electromagnetic dipole emission. Here we have analyzed the regime of spin down losses  dominated by GW emission, that applies when $\epsilon_{-5}\gtrsim 5\  P_{-3} B_{p,12}$. In this case
 the X-ray luminosity is much lower than the EM-dominated regime, by a factor $\approx t/\tau_*$ thus relaxing the constraints on the dipole magnetic field (see also \citealt{ai18}).
 By requiring that the expected flux be below the observed flux we derive a joint constraint on the ellipticity and the dipolar component of the magnetic field, $\epsilon_{-5}\gtrsim 5 \eta_{-3} B_{p,12}$. 
 
 Such an ellipticity can be produced by the strong toroidal field that develops  due to the differential rotation in a nascent NS after the merger \citep{rezzolla18,giacomazzo15}. The strong toroidal field can be responsible for the candidate X-ray flare detected, at a $\gtrsim 3  \sigma$ significance, 155 days after the merger. Indeed various properties of the flare (relative duration and amplitude, luminosity, curvature effect) are consistent with those observed in X-ray flares, and attributed to a long-lived central engine. We  argue that this could also be  the case for GW170817, specifically calling for a long-lived magnetized NS characterized by a strong toroidal component.  
The untwisting of the toroidal magnetic field may give rise to an abrupt injection of outflows \citep{thompson01,dai06}, and the internal magnetic dissipation of such an outflow \citep{zhangyan11} would give rise to the temporal variability observable in X-rays. 
From the total energy in the flare we estimate that the toroidal component of the magnetic field has to be $B_{\rm t} \gtrsim 10^{14}$ G, which is reasonably expected \citep{thompson93}.

In conclusion, our model  envisions a structured jet launched by rapidly spinning long-lived NS with a strong differential magnetic field (Figure~\ref{fig:cartoon}). The structured jet fully accounts for the broad band non-thermal continuum. 
The existence of a central engine pulsar would inevitably provide additional energy injection to the blastwave and to the kilonova ejecta. This would influence the emission properties of the broad-band afterglow and the kilonova emission.
The impact on the kilonova due to the energy injection of the underlying pulsar has been studied \citep{yu17,li18}. Both the early (blue) and late (red) kilonova components can be accounted for with reasonable values of ejected mass and opacity if the neutron star spindown is dominated by gravitational wave losses \citep{li18}. In this regime we  have verified that energy injection into the blast-wave due to central engine is negligibly small, which does not affect the best fitting parameters of the structured jet. While energy injection from the pulsar has a negligible effect on the observed afterglow, it may cause a flattening  at late times  ($\gtrsim 2$ yr). Abrupt magnetic reconnection of the strong toroidal component  launches brief relativistic outflow that produce X-ray flares via magnetic dissipation.

The sparse sampling of the afterglow did not allow us to robustly detect and characterize its temporal variability. Future X-ray campaigns of GW counterparts should aim at  providing adequate sampling of the light curve, needed to firmly establish and characterize  short-term temporal variability  and its connection with the central engine.

If the remnant of GW170817 is a long-lived NS, then the maximum mass 
of a non-spinning NS should be at least greater than $2.16 M_\odot$ \citep{ruiz18, rezzolla18,margalit17}, superseding the current lower limit of $2 M_\odot$ set by PSR J1614-2230 \citep{demorest10}. This new limit would eliminate essentially all the soft neutron star equations of state invoking hyperons and boson condensation \citep{lattimer07} and would support the suggestion \citep{gao16} that a good fraction of NS-NS mergers leave behind supra-massive or stable NSs.

\section*{Appendix: Origin of the X-ray variability: afterglow}

The rapid variability $\Delta t / t \lesssim 0.15 $ places our candidate flare in a region excluded by afterglow models \citep[Figure~\ref{fig:ioka}]{ioka05,Burrows05,Piro05}. 
At 160~d the forward shock is still moving at a mildly relativistic velocity.  The light crossing time across the shock front is then of the same order as the time since the explosion, i.e. $\Delta t \approx t$ \citep{Kumar&Piran00}, much 
longer than observed.  In principle a small region of angular size $\Delta \theta$ such that $\Delta t \gtrsim R \Delta \theta {\rm max}(\Delta \theta/2, 2\theta_v)/c$ can accomodate the observed timescale \citep{ioka05}.  However, it has been demonstrated both analytically and numerically that, even for strong density perturbations, flux changes are smoothed over much longer time scales \citep{NakarGranot2007, GatvanEertenMacFadyen2013,uhm14}.
A further argument is the following.
By taking into account the volume of the variable region and the volume of the observable region one derives an upper limit 
\begin{eqnarray}
\Delta F_\nu/F_\nu
\lesssim \left\{
\begin{array}{ll}
4/5\  \Delta t/t\  f_{\rm enhance}& ({\rm on-axis})\\
6 (\Delta t/t)^2\ f_{\rm enhance}& ({\rm off-axis})
\end{array}
\right.
\label{eq:ioka1}
\end{eqnarray}
where the enhancement due to a overdensity $n_{f}$ is $f_{\rm enhance}=(\nu_{c,f}/\nu_c)^{-1/2}-1=(n_{f}/n)^{1/2}-1$, where $\nu_{c,f}$ is  the cooling frequency  of the blob. 
When the density increases as much as to shift the cooling frequency below the observed frequency, there is no longer a gain and the flux remains constant. Thus the maximum gain is $f_{\rm enhance}\approx(\nu_c/\nu_x)^{1/2}$. 
From equation~\ref{eq:ioka1}, in order to satisfy the flare properties requires $\nu_c\gtrsim 10^{21}$ Hz. This is not consistent with the value derived for the structured jet model and would require an unplausible  low density of the ISM  $n\lesssim 10^{-7}$ cm$^{-3}$ for the cocoon model.
At the projected distance of GW170817, massive elliptical and S0 galaxies typically have particle densities of $\approx 10^{-2}$ cm$^{-3}$ \citep{Lakhchaura18}.  Even accounting for the smaller mass of NGC4993 (about a factor of four smaller than the median of the
Lakhchaura sample), this is still orders of magnitude larger than required for the cocoon model.

In the case of a cocoon, where energy injection by an outflow with a spread of Lorentz factors drives the shock,  a strong modulation of the profile over the assumed power-law can produce a bump in the light curve when e.g. a  massive late relativistic shell catches up with the shock front.  However  this interaction  will produce bumps that have typically $\Delta t \approx t$ \citep{kumar00}, thus much longer than observed. In addition the predicted stepwise increase above the baseline does not reproduce the observed flare-like feature. 
 In the case of a structured jet while the broader and slower component will quickly lose its energy in the environment, the (faster) narrow-core of the jet will excavate a free path to the slower ejecta in its wave, thus allowing   $\Delta t \ll t$ \citep{Granot+03}. However, as in the previous case, a stepwise light curve is expected.
Finally, a structured jet with a significant angular structure (patchy jet) would also give a similar variability time scale $\Delta t \approx t$, and therefore disfavored.


\section*{Acknowledgements}
LP acknowledges partial support by  the
European Union Horizon 2020 Programme under the AHEAD project (grant
agreement number 654215). 




\bibliographystyle{mnras}

\bibliography{grb_ref} 

\begin{thebibliography}{}
\makeatletter
\relax
\def\mn@urlcharsother{\let\do\@makeother \do\$\do\&\do\#\do\^\do\_\do\%\do\~}
\def\mn@doi{\begingroup\mn@urlcharsother \@ifnextchar [ {\mn@doi@}
  {\mn@doi@[]}}
\def\mn@doi@[#1]#2{\def\@tempa{#1}\ifx\@tempa\@empty \href
  {http://dx.doi.org/#2} {doi:#2}\else \href {http://dx.doi.org/#2} {#1}\fi
  \endgroup}
\def\mn@eprint#1#2{\mn@eprint@#1:#2::\@nil}
\def\mn@eprint@arXiv#1{\href {http://arxiv.org/abs/#1} {{\tt arXiv:#1}}}
\def\mn@eprint@dblp#1{\href {http://dblp.uni-trier.de/rec/bibtex/#1.xml}
  {dblp:#1}}
\def\mn@eprint@#1:#2:#3:#4\@nil{\def\@tempa {#1}\def\@tempb {#2}\def\@tempc
  {#3}\ifx \@tempc \@empty \let \@tempc \@tempb \let \@tempb \@tempa \fi \ifx
  \@tempb \@empty \def\@tempb {arXiv}\fi \@ifundefined
  {mn@eprint@\@tempb}{\@tempb:\@tempc}{\expandafter \expandafter \csname
  mn@eprint@\@tempb\endcsname \expandafter{\@tempc}}}

\bibitem[\protect\citeauthoryear{{Abbott} et~al.,}{{Abbott}
  et~al.}{2017a}]{GW170817-multi}
{Abbott} B.~P.,  et~al., 2017a, \mn@doi [\apjl] {10.3847/2041-8213/aa91c9},
  \href {http://adsabs.harvard.edu/abs/2017ApJ...848L..12A} {848, L12}

\bibitem[\protect\citeauthoryear{{Abbott} et~al.,}{{Abbott}
  et~al.}{2017b}]{LIGO-product}
{Abbott} B.~P.,  et~al., 2017b, \mn@doi [\apjl] {10.3847/2041-8213/aa9a35},
  \href {http://adsabs.harvard.edu/abs/2017ApJ...851L..16A} {851, L16}

\bibitem[\protect\citeauthoryear{{Ai}, {Gao}, {Dai}, {Wu}, {Li}, {Zhang}  \&
  {Li}}{{Ai} et~al.}{2018}]{ai18}
{Ai} S.,  {Gao} H.,  {Dai} Z.-G.,  {Wu} X.-F.,  {Li} A.,  {Zhang} B.,   {Li}
  M.-Z.,  2018, \mn@doi [\apj] {10.3847/1538-4357/aac2b7}, \href
  {http://adsabs.harvard.edu/abs/2018ApJ...860...57A} {860, 57}

\bibitem[\protect\citeauthoryear{{Aloy}, {Janka}  \& {M{\"u}ller}}{{Aloy}
  et~al.}{2005}]{AloyJankaMueller2005}
{Aloy} M.~A.,  {Janka} H.-T.,   {M{\"u}ller} E.,  2005, \mn@doi [\aap]
  {10.1051/0004-6361:20041865}, \href
  {http://adsabs.harvard.edu/abs/2005A%26A...436..273A} {436, 273}

\bibitem[\protect\citeauthoryear{{Bernardini}, {Margutti}, {Chincarini},
  {Guidorzi}  \& {Mao}}{{Bernardini} et~al.}{2011}]{Bernardini2011}
{Bernardini} M.~G.,  {Margutti} R.,  {Chincarini} G.,  {Guidorzi} C.,   {Mao}
  J.,  2011, \mn@doi [\aap] {10.1051/0004-6361/201015703}, \href
  {http://adsabs.harvard.edu/abs/2011A%26A...526A..27B} {526, A27}

\bibitem[\protect\citeauthoryear{{Breu} \& {Rezzolla}}{{Breu} \&
  {Rezzolla}}{2016}]{breu16}
{Breu} C.,  {Rezzolla} L.,  2016, \mn@doi [\mnras] {10.1093/mnras/stw575},
  \href {http://adsabs.harvard.edu/abs/2016MNRAS.459..646B} {459, 646}

\bibitem[\protect\citeauthoryear{{Burrows} et~al.,}{{Burrows}
  et~al.}{2005}]{Burrows05}
{Burrows} D.~N.,  et~al., 2005, \mn@doi [Science] {10.1126/science.1116168},
  \href {http://adsabs.harvard.edu/abs/2005Sci...309.1833B} {309, 1833}

\bibitem[\protect\citeauthoryear{{Cash}}{{Cash}}{1979}]{Cash79}
{Cash} W.,  1979, \mn@doi [\apj] {10.1086/156922}, \href
  {http://adsabs.harvard.edu/abs/1979ApJ...228..939C} {228, 939}

\bibitem[\protect\citeauthoryear{{Chincarini} et~al.,}{{Chincarini}
  et~al.}{2007}]{Chincarini07}
{Chincarini} G.,  et~al., 2007, \mn@doi [\apj] {10.1086/521591}, \href
  {http://adsabs.harvard.edu/abs/2007ApJ...671.1903C} {671, 1903}

\bibitem[\protect\citeauthoryear{{Corsi} \& {M{\'e}sz{\'a}ros}}{{Corsi} \&
  {M{\'e}sz{\'a}ros}}{2009}]{Corsi09}
{Corsi} A.,  {M{\'e}sz{\'a}ros} P.,  2009, \mn@doi [\apj]
  {10.1088/0004-637X/702/2/1171}, \href
  {http://adsabs.harvard.edu/abs/2009ApJ...702.1171C} {702, 1171}

\bibitem[\protect\citeauthoryear{{Cutler}}{{Cutler}}{2002}]{cutler02}
{Cutler} C.,  2002, \mn@doi [\prd] {10.1103/PhysRevD.66.084025}, \href
  {http://adsabs.harvard.edu/abs/2002PhRvD..66h4025C} {66, 084025}

\bibitem[\protect\citeauthoryear{{D'Avanzo} et~al.,}{{D'Avanzo}
  et~al.}{2018}]{Davanzo2018}
{D'Avanzo} P.,  et~al., 2018, preprint, \href
  {http://adsabs.harvard.edu/abs/2018arXiv180106164D} {} (\mn@eprint {arXiv}
  {1801.06164})

\bibitem[\protect\citeauthoryear{{Dai} \& {Lu}}{{Dai} \& {Lu}}{1998}]{dai98}
{Dai} Z.~G.,  {Lu} T.,  1998, \aap, \href
  {http://adsabs.harvard.edu/abs/1998A%26A...333L..87D} {333, L87}

\bibitem[\protect\citeauthoryear{{Dai}, {Wang}, {Wu}  \& {Zhang}}{{Dai}
  et~al.}{2006}]{dai06}
{Dai} Z.~G.,  {Wang} X.~Y.,  {Wu} X.~F.,   {Zhang} B.,  2006, \mn@doi [Science]
  {10.1126/science.1123606}, \href
  {http://adsabs.harvard.edu/abs/2006Sci...311.1127D} {311, 1127}

\bibitem[\protect\citeauthoryear{{Dall'Osso} \& {Stella}}{{Dall'Osso} \&
  {Stella}}{2007}]{dallosso07}
{Dall'Osso} S.,  {Stella} L.,  2007, \mn@doi [\apss]
  {10.1007/s10509-007-9323-0}, \href
  {http://adsabs.harvard.edu/abs/2007Ap%26SS.308..119D} {308, 119}

\bibitem[\protect\citeauthoryear{{Dall'Osso}, {Shore}  \& {Stella}}{{Dall'Osso}
  et~al.}{2009}]{dallosso09}
{Dall'Osso} S.,  {Shore} S.~N.,   {Stella} L.,  2009, \mn@doi [\mnras]
  {10.1111/j.1365-2966.2008.14054.x}, \href
  {http://adsabs.harvard.edu/abs/2009MNRAS.398.1869D} {398, 1869}

\bibitem[\protect\citeauthoryear{{Dall'Osso}, {Stella}  \&
  {Palomba}}{{Dall'Osso} et~al.}{2018}]{dallosso18}
{Dall'Osso} S.,  {Stella} L.,   {Palomba} C.,  2018, \mn@doi [\mnras]
  {10.1093/mnras/sty1706}, \href
  {http://adsabs.harvard.edu/abs/2018MNRAS.480.1353D} {480, 1353}

\bibitem[\protect\citeauthoryear{{Demorest}, {Pennucci}, {Ransom}, {Roberts}
  \& {Hessels}}{{Demorest} et~al.}{2010}]{demorest10}
{Demorest} P.~B.,  {Pennucci} T.,  {Ransom} S.~M.,  {Roberts} M.~S.~E.,
  {Hessels} J.~W.~T.,  2010, \mn@doi [\nat] {10.1038/nature09466}, \href
  {http://adsabs.harvard.edu/abs/2010Natur.467.1081D} {467, 1081}

\bibitem[\protect\citeauthoryear{{Deng}, {Li}, {Zhang}  \& {Li}}{{Deng}
  et~al.}{2015}]{deng15}
{Deng} W.,  {Li} H.,  {Zhang} B.,   {Li} S.,  2015, \mn@doi [\apj]
  {10.1088/0004-637X/805/2/163}, \href
  {http://adsabs.harvard.edu/abs/2015ApJ...805..163D} {805, 163}

\bibitem[\protect\citeauthoryear{{Deustua}}{{Deustua}}{2016}]{hst01}
{Deustua} S.,  2016, Space Telescope Science Institute

\bibitem[\protect\citeauthoryear{{Dobie} et~al.,}{{Dobie}
  et~al.}{2018}]{Dobie18}
{Dobie} D.,  et~al., 2018, \mn@doi [\apjl] {10.3847/2041-8213/aac105}, \href
  {http://adsabs.harvard.edu/abs/2018ApJ...858L..15D} {858, L15}

\bibitem[\protect\citeauthoryear{{Evans} et~al.,}{{Evans}
  et~al.}{2017}]{SwiftGW}
{Evans} P.~A.,  et~al., 2017, \mn@doi [Science] {10.1126/science.aap9580},
  \href {http://adsabs.harvard.edu/abs/2017Sci...358.1565E} {358, 1565}

\bibitem[\protect\citeauthoryear{{Fan}, {Wu}  \& {Wei}}{{Fan}
  et~al.}{2013}]{fan13}
{Fan} Y.-Z.,  {Wu} X.-F.,   {Wei} D.-M.,  2013, \mn@doi [\prd]
  {10.1103/PhysRevD.88.067304}, \href
  {http://adsabs.harvard.edu/abs/2013PhRvD..88f7304F} {88, 067304}

\bibitem[\protect\citeauthoryear{{Foreman-Mackey}, {Hogg}, {Lang}  \&
  {Goodman}}{{Foreman-Mackey} et~al.}{2013}]{Foreman-Mackey2013}
{Foreman-Mackey} D.,  {Hogg} D.~W.,  {Lang} D.,   {Goodman} J.,  2013, \mn@doi
  [\pasp] {10.1086/670067}, \href
  {http://adsabs.harvard.edu/abs/2013PASP..125..306F} {125, 306}

\bibitem[\protect\citeauthoryear{{Gao}, {Ding}, {Wu}, {Dai}  \& {Zhang}}{{Gao}
  et~al.}{2015}]{gao15}
{Gao} H.,  {Ding} X.,  {Wu} X.-F.,  {Dai} Z.-G.,   {Zhang} B.,  2015, \mn@doi
  [\apj] {10.1088/0004-637X/807/2/163}, \href
  {http://adsabs.harvard.edu/abs/2015ApJ...807..163G} {807, 163}

\bibitem[\protect\citeauthoryear{{Gao}, {Zhang}  \& {L{\"u}}}{{Gao}
  et~al.}{2016}]{gao16}
{Gao} H.,  {Zhang} B.,   {L{\"u}} H.-J.,  2016, \mn@doi [\prd]
  {10.1103/PhysRevD.93.044065}, \href
  {http://adsabs.harvard.edu/abs/2016PhRvD..93d4065G} {93, 044065}

\bibitem[\protect\citeauthoryear{{Gat}, {van Eerten}  \& {MacFadyen}}{{Gat}
  et~al.}{2013}]{GatvanEertenMacFadyen2013}
{Gat} I.,  {van Eerten} H.,   {MacFadyen} A.,  2013, \mn@doi [\apj]
  {10.1088/0004-637X/773/1/2}, \href
  {http://adsabs.harvard.edu/abs/2013ApJ...773....2G} {773, 2}

\bibitem[\protect\citeauthoryear{{Ghirlanda} et~al.,}{{Ghirlanda}
  et~al.}{2018}]{Ghirlanda18}
{Ghirlanda} G.,  et~al., 2018, preprint, \href
  {http://adsabs.harvard.edu/abs/2018arXiv180800469G} {} (\mn@eprint {arXiv}
  {1808.00469})

\bibitem[\protect\citeauthoryear{{Giacomazzo}, {Zrake}, {Duffell}, {MacFadyen}
  \& {Perna}}{{Giacomazzo} et~al.}{2015}]{giacomazzo15}
{Giacomazzo} B.,  {Zrake} J.,  {Duffell} P.~C.,  {MacFadyen} A.~I.,   {Perna}
  R.,  2015, \mn@doi [\apj] {10.1088/0004-637X/809/1/39}, \href
  {http://adsabs.harvard.edu/abs/2015ApJ...809...39G} {809, 39}

\bibitem[\protect\citeauthoryear{{Goldstein} et~al.,}{{Goldstein}
  et~al.}{2017}]{goldstein17}
{Goldstein} A.,  et~al., 2017, \mn@doi [\apjl] {10.3847/2041-8213/aa8f41},
  \href {http://adsabs.harvard.edu/abs/2017ApJ...848L..14G} {848, L14}

\bibitem[\protect\citeauthoryear{{Gonzaga}, {Hack}, {Fruchter}  \&
  {Mack}}{{Gonzaga} et~al.}{2012}]{hst02}
{Gonzaga} S.,  {Hack} W.,  {Fruchter} A.,   {Mack} J.,  2012, {The DrizzlePac
  Handbook}

\bibitem[\protect\citeauthoryear{{Granot}, {Nakar}  \& {Piran}}{{Granot}
  et~al.}{2003}]{Granot+03}
{Granot} J.,  {Nakar} E.,   {Piran} T.,  2003, \mn@doi [\nat]
  {10.1038/426138a}, \href {http://adsabs.harvard.edu/abs/2003Natur.426..138G}
  {426, 138}

\bibitem[\protect\citeauthoryear{{Haggard}, {Nynka}, {Ruan}, {Kalogera},
  {Cenko}, {Evans}  \& {Kennea}}{{Haggard} et~al.}{2017}]{haggard2017}
{Haggard} D.,  {Nynka} M.,  {Ruan} J.~J.,  {Kalogera} V.,  {Cenko} S.~B.,
  {Evans} P.,   {Kennea} J.~A.,  2017, \mn@doi [\apjl]
  {10.3847/2041-8213/aa8ede}, \href
  {http://adsabs.harvard.edu/abs/2017ApJ...848L..25H} {848, L25}

\bibitem[\protect\citeauthoryear{{Ioka}, {Kobayashi}  \& {Zhang}}{{Ioka}
  et~al.}{2005}]{ioka05}
{Ioka} K.,  {Kobayashi} S.,   {Zhang} B.,  2005, \mn@doi [\apj]
  {10.1086/432567}, \href {http://adsabs.harvard.edu/abs/2005ApJ...631..429I}
  {631, 429}

\bibitem[\protect\citeauthoryear{{Jones}}{{Jones}}{1976}]{jones76}
{Jones} P.~B.,  1976, \mn@doi [\apss] {10.1007/BF00642671}, \href
  {http://adsabs.harvard.edu/abs/1976Ap%26SS..45..369J} {45, 369}

\bibitem[\protect\citeauthoryear{{Kargaltsev}, {Rangelov}  \&
  {Pavlov}}{{Kargaltsev} et~al.}{2013}]{kargaltsev13}
{Kargaltsev} O.,  {Rangelov} B.,   {Pavlov} G.~G.,  2013, preprint, \href
  {http://adsabs.harvard.edu/abs/2013arXiv1305.2552K} {} (\mn@eprint {arXiv}
  {1305.2552})

\bibitem[\protect\citeauthoryear{{Kasen}, {Fern{\'a}ndez}  \&
  {Metzger}}{{Kasen} et~al.}{2015}]{kasen15}
{Kasen} D.,  {Fern{\'a}ndez} R.,   {Metzger} B.~D.,  2015, \mn@doi [\mnras]
  {10.1093/mnras/stv721}, \href
  {http://adsabs.harvard.edu/abs/2015MNRAS.450.1777K} {450, 1777}

\bibitem[\protect\citeauthoryear{{Kasen}, {Metzger}, {Barnes}, {Quataert}  \&
  {Ramirez-Ruiz}}{{Kasen} et~al.}{2017}]{kasen17}
{Kasen} D.,  {Metzger} B.,  {Barnes} J.,  {Quataert} E.,   {Ramirez-Ruiz} E.,
  2017, \mn@doi [\nat] {10.1038/nature24453}, \href
  {http://adsabs.harvard.edu/abs/2017Natur.551...80K} {551, 80}

\bibitem[\protect\citeauthoryear{{Kathirgamaraju}, {Barniol Duran}  \&
  {Giannios}}{{Kathirgamaraju} et~al.}{2018}]{kathi18}
{Kathirgamaraju} A.,  {Barniol Duran} R.,   {Giannios} D.,  2018, \mn@doi
  [\mnras] {10.1093/mnrasl/slx175}, \href
  {http://adsabs.harvard.edu/abs/2018MNRAS.473L.121K} {473, L121}

\bibitem[\protect\citeauthoryear{{Kumar} \& {Panaitescu}}{{Kumar} \&
  {Panaitescu}}{2000}]{kumar00}
{Kumar} P.,  {Panaitescu} A.,  2000, \mn@doi [\apjl] {10.1086/312905}, \href
  {http://adsabs.harvard.edu/abs/2000ApJ...541L..51K} {541, L51}

\bibitem[\protect\citeauthoryear{{Kumar} \& {Piran}}{{Kumar} \&
  {Piran}}{2000}]{Kumar&Piran00}
{Kumar} P.,  {Piran} T.,  2000, \mn@doi [\apj] {10.1086/308537}, \href
  {http://adsabs.harvard.edu/abs/2000ApJ...532..286K} {532, 286}

\bibitem[\protect\citeauthoryear{{Lakhchaura} et~al.,}{{Lakhchaura}
  et~al.}{2018}]{Lakhchaura18}
{Lakhchaura} K.,  et~al., 2018, \mn@doi [\mnras] {10.1093/mnras/sty2565}, \href
  {http://adsabs.harvard.edu/abs/2018MNRAS.481.4472L} {481, 4472}

\bibitem[\protect\citeauthoryear{{Lasker} et~al.,}{{Lasker}
  et~al.}{2008}]{Lasker08}
{Lasker} B.~M.,  et~al., 2008, \mn@doi [\aj] {10.1088/0004-6256/136/2/735},
  \href {http://adsabs.harvard.edu/abs/2008AJ....136..735L} {136, 735}

\bibitem[\protect\citeauthoryear{{Lasky} \& {Glampedakis}}{{Lasky} \&
  {Glampedakis}}{2016}]{Lasky16}
{Lasky} P.~D.,  {Glampedakis} K.,  2016, \mn@doi [\mnras]
  {10.1093/mnras/stw435}, \href
  {http://adsabs.harvard.edu/abs/2016MNRAS.458.1660L} {458, 1660}

\bibitem[\protect\citeauthoryear{{Lattimer} \& {Prakash}}{{Lattimer} \&
  {Prakash}}{2007}]{lattimer07}
{Lattimer} J.~M.,  {Prakash} M.,  2007, \mn@doi [\physrep]
  {10.1016/j.physrep.2007.02.003}, \href
  {http://adsabs.harvard.edu/abs/2007PhR...442..109L} {442, 109}

\bibitem[\protect\citeauthoryear{{Lazarian}, {Zhang}  \& {Xu}}{{Lazarian}
  et~al.}{2018}]{lazarian18}
{Lazarian} A.,  {Zhang} B.,   {Xu} S.,  2018, preprint, \href
  {http://adsabs.harvard.edu/abs/2018arXiv180104061L} {} (\mn@eprint {arXiv}
  {1801.04061})

\bibitem[\protect\citeauthoryear{{Lazzati}, {L{\'o}pez-C{\'a}mara},
  {Cantiello}, {Morsony}, {Perna}  \& {Workman}}{{Lazzati}
  et~al.}{2017}]{Lazzati2017}
{Lazzati} D.,  {L{\'o}pez-C{\'a}mara} D.,  {Cantiello} M.,  {Morsony} B.~J.,
  {Perna} R.,   {Workman} J.~C.,  2017, \mn@doi [\apjl]
  {10.3847/2041-8213/aa8f3d}, \href
  {http://adsabs.harvard.edu/abs/2017ApJ...848L...6L} {848, L6}

\bibitem[\protect\citeauthoryear{{Lazzati}, {Perna}, {Morsony}, {Lopez-Camara},
  {Cantiello}, {Ciolfi}, {Giacomazzo}  \& {Workman}}{{Lazzati}
  et~al.}{2018}]{Lazzati2018}
{Lazzati} D.,  {Perna} R.,  {Morsony} B.~J.,  {Lopez-Camara} D.,  {Cantiello}
  M.,  {Ciolfi} R.,  {Giacomazzo} B.,   {Workman} J.~C.,  2018, \mn@doi
  [Physical Review Letters] {10.1103/PhysRevLett.120.241103}, \href
  {http://adsabs.harvard.edu/abs/2018PhRvL.120x1103L} {120, 241103}

\bibitem[\protect\citeauthoryear{{Li}, {Liu}, {Yu}  \& {Zhang}}{{Li}
  et~al.}{2018}]{li18}
{Li} S.-Z.,  {Liu} L.-D.,  {Yu} Y.-W.,   {Zhang} B.,  2018, \mn@doi [\apjl]
  {10.3847/2041-8213/aace61}, \href
  {http://adsabs.harvard.edu/abs/2018ApJ...861L..12L} {861, L12}

\bibitem[\protect\citeauthoryear{{Liang} et~al.,}{{Liang}
  et~al.}{2006}]{liang06}
{Liang} E.~W.,  et~al., 2006, \mn@doi [\apj] {10.1086/504684}, \href
  {http://adsabs.harvard.edu/abs/2006ApJ...646..351L} {646, 351}

\bibitem[\protect\citeauthoryear{{L{\"u}}, {Zhang}, {Lei}, {Li}  \&
  {Lasky}}{{L{\"u}} et~al.}{2015}]{lv15}
{L{\"u}} H.-J.,  {Zhang} B.,  {Lei} W.-H.,  {Li} Y.,   {Lasky} P.~D.,  2015,
  \mn@doi [\apj] {10.1088/0004-637X/805/2/89}, \href
  {http://adsabs.harvard.edu/abs/2015ApJ...805...89L} {805, 89}

\bibitem[\protect\citeauthoryear{{Lyman} et~al.,}{{Lyman}
  et~al.}{2018}]{Lyman18}
{Lyman} J.~D.,  et~al., 2018, preprint, \href
  {http://adsabs.harvard.edu/abs/2018arXiv180102669L} {} (\mn@eprint {arXiv}
  {1801.02669})

\bibitem[\protect\citeauthoryear{{Margalit} \& {Metzger}}{{Margalit} \&
  {Metzger}}{2017}]{margalit17}
{Margalit} B.,  {Metzger} B.~D.,  2017, \mn@doi [\apjl]
  {10.3847/2041-8213/aa991c}, \href
  {http://adsabs.harvard.edu/abs/2017ApJ...850L..19M} {850, L19}

\bibitem[\protect\citeauthoryear{{Margutti} et~al.,}{{Margutti}
  et~al.}{2018}]{Margutti18}
{Margutti} R.,  et~al., 2018, \mn@doi [\apjl] {10.3847/2041-8213/aab2ad}, \href
  {http://adsabs.harvard.edu/abs/2018ApJ...856L..18M} {856, L18}

\bibitem[\protect\citeauthoryear{{Metzger} \& {Piro}}{{Metzger} \&
  {Piro}}{2014}]{metzger14}
{Metzger} B.~D.,  {Piro} A.~L.,  2014, \mn@doi [\mnras] {10.1093/mnras/stu247},
  \href {http://adsabs.harvard.edu/abs/2014MNRAS.439.3916M} {439, 3916}

\bibitem[\protect\citeauthoryear{{Metzger}, {Quataert}  \&
  {Thompson}}{{Metzger} et~al.}{2008}]{metzger08}
{Metzger} B.~D.,  {Quataert} E.,   {Thompson} T.~A.,  2008, \mn@doi [\mnras]
  {10.1111/j.1365-2966.2008.12923.x}, \href
  {http://adsabs.harvard.edu/abs/2008MNRAS.385.1455M} {385, 1455}

\bibitem[\protect\citeauthoryear{{Metzger}, {Thompson}  \&
  {Quataert}}{{Metzger} et~al.}{2018}]{metzger18}
{Metzger} B.~D.,  {Thompson} T.~A.,   {Quataert} E.,  2018, \mn@doi [\apj]
  {10.3847/1538-4357/aab095}, \href
  {http://adsabs.harvard.edu/abs/2018ApJ...856..101M} {856, 101}

\bibitem[\protect\citeauthoryear{{Mooley} et~al.,}{{Mooley}
  et~al.}{2018a}]{Mooley2018}
{Mooley} K.~P.,  et~al., 2018a, \mn@doi [\nat] {10.1038/nature25452}, \href
  {http://adsabs.harvard.edu/abs/2018Natur.554..207M} {554, 207}

\bibitem[\protect\citeauthoryear{{Mooley} et~al.,}{{Mooley}
  et~al.}{2018b}]{Mooley18b}
{Mooley} K.~P.,  et~al., 2018b, \mn@doi [\nat] {10.1038/s41586-018-0486-3},
  \href {http://adsabs.harvard.edu/abs/2018Natur.561..355M} {561, 355}

\bibitem[\protect\citeauthoryear{{Murase} et~al.,}{{Murase}
  et~al.}{2018}]{murase18}
{Murase} K.,  et~al., 2018, \mn@doi [\apj] {10.3847/1538-4357/aaa48a}, \href
  {http://adsabs.harvard.edu/abs/2018ApJ...854...60M} {854, 60}

\bibitem[\protect\citeauthoryear{{Nakar} \& {Granot}}{{Nakar} \&
  {Granot}}{2007}]{NakarGranot2007}
{Nakar} E.,  {Granot} J.,  2007, \mn@doi [\mnras]
  {10.1111/j.1365-2966.2007.12245.x}, \href
  {http://adsabs.harvard.edu/abs/2007MNRAS.380.1744N} {380, 1744}

\bibitem[\protect\citeauthoryear{{Perna}, {Armitage}  \& {Zhang}}{{Perna}
  et~al.}{2006}]{perna06}
{Perna} R.,  {Armitage} P.~J.,   {Zhang} B.,  2006, \mn@doi [\apjl]
  {10.1086/499775}, \href {http://adsabs.harvard.edu/abs/2006ApJ...636L..29P}
  {636, L29}

\bibitem[\protect\citeauthoryear{{Pian} et~al.,}{{Pian} et~al.}{2017}]{Pian17}
{Pian} E.,  et~al., 2017, \mn@doi [\nat] {10.1038/nature24298}, \href
  {http://adsabs.harvard.edu/abs/2017Natur.551...67P} {551, 67}

\bibitem[\protect\citeauthoryear{{Piro} et~al.,}{{Piro} et~al.}{2005}]{Piro05}
{Piro} L.,  et~al., 2005, \mn@doi [\apj] {10.1086/428377}, \href
  {http://adsabs.harvard.edu/abs/2005ApJ...623..314P} {623, 314}

\bibitem[\protect\citeauthoryear{{Pooley}, {Kumar}, {Wheeler}  \&
  {Grossan}}{{Pooley} et~al.}{2018}]{pooley18}
{Pooley} D.,  {Kumar} P.,  {Wheeler} J.~C.,   {Grossan} B.,  2018, \mn@doi
  [\apjl] {10.3847/2041-8213/aac3d6}, \href
  {http://adsabs.harvard.edu/abs/2018ApJ...859L..23P} {859, L23}

\bibitem[\protect\citeauthoryear{{Radice}, {Perego}, {Bernuzzi}  \&
  {Zhang}}{{Radice} et~al.}{2018}]{radice18}
{Radice} D.,  {Perego} A.,  {Bernuzzi} S.,   {Zhang} B.,  2018, \mn@doi
  [\mnras] {10.1093/mnras/sty2531}, \href
  {http://adsabs.harvard.edu/abs/2018MNRAS.481.3670R} {481, 3670}

\bibitem[\protect\citeauthoryear{{Rea} et~al.,}{{Rea} et~al.}{2010}]{rea10}
{Rea} N.,  et~al., 2010, \mn@doi [Science] {10.1126/science.1196088}, \href
  {http://adsabs.harvard.edu/abs/2010Sci...330..944R} {330, 944}

\bibitem[\protect\citeauthoryear{{Rezzolla}, {Most}  \& {Weih}}{{Rezzolla}
  et~al.}{2018}]{rezzolla18}
{Rezzolla} L.,  {Most} E.~R.,   {Weih} L.~R.,  2018, \mn@doi [\apjl]
  {10.3847/2041-8213/aaa401}, \href
  {http://adsabs.harvard.edu/abs/2018ApJ...852L..25R} {852, L25}

\bibitem[\protect\citeauthoryear{{Rosswog}}{{Rosswog}}{2007}]{Rosswog2007}
{Rosswog} S.,  2007, \mn@doi [\mnras] {10.1111/j.1745-3933.2007.00284.x}, \href
  {http://adsabs.harvard.edu/abs/2007MNRAS.376L..48R} {376, L48}

\bibitem[\protect\citeauthoryear{{Rowlinson}, {O'Brien}, {Metzger}, {Tanvir}
  \& {Levan}}{{Rowlinson} et~al.}{2013}]{rowlinson13}
{Rowlinson} A.,  {O'Brien} P.~T.,  {Metzger} B.~D.,  {Tanvir} N.~R.,   {Levan}
  A.~J.,  2013, \mn@doi [\mnras] {10.1093/mnras/sts683}, \href
  {http://adsabs.harvard.edu/abs/2013MNRAS.430.1061R} {430, 1061}

\bibitem[\protect\citeauthoryear{{Ruiz}, {Shapiro}  \& {Tsokaros}}{{Ruiz}
  et~al.}{2018}]{ruiz18}
{Ruiz} M.,  {Shapiro} S.~L.,   {Tsokaros} A.,  2018, \mn@doi [\prd]
  {10.1103/PhysRevD.97.021501}, \href
  {http://adsabs.harvard.edu/abs/2018PhRvD..97b1501R} {97, 021501}

\bibitem[\protect\citeauthoryear{{Savchenko} et~al.,}{{Savchenko}
  et~al.}{2017}]{Savchenko17}
{Savchenko} V.,  et~al., 2017, \mn@doi [\apjl] {10.3847/2041-8213/aa8f94},
  \href {http://adsabs.harvard.edu/abs/2017ApJ...848L..15S} {848, L15}

\bibitem[\protect\citeauthoryear{{Schlafly} \& {Finkbeiner}}{{Schlafly} \&
  {Finkbeiner}}{2011}]{sf11}
{Schlafly} E.~F.,  {Finkbeiner} D.~P.,  2011, \mn@doi [\apj]
  {10.1088/0004-637X/737/2/103}, \href
  {http://adsabs.harvard.edu/abs/2011ApJ...737..103S} {737, 103}

\bibitem[\protect\citeauthoryear{{Smartt} et~al.,}{{Smartt}
  et~al.}{2017}]{Smartt17}
{Smartt} S.~J.,  et~al., 2017, \mn@doi [\nat] {10.1038/nature24303}, \href
  {http://adsabs.harvard.edu/abs/2017Natur.551...75S} {551, 75}

\bibitem[\protect\citeauthoryear{{Tanvir} et~al.,}{{Tanvir}
  et~al.}{2017}]{tanvir17}
{Tanvir} N.~R.,  et~al., 2017, \mn@doi [\apjl] {10.3847/2041-8213/aa90b6},
  \href {http://adsabs.harvard.edu/abs/2017ApJ...848L..27T} {848, L27}

\bibitem[\protect\citeauthoryear{{Thompson} \& {Duncan}}{{Thompson} \&
  {Duncan}}{1993}]{thompson93}
{Thompson} C.,  {Duncan} R.~C.,  1993, \mn@doi [\apj] {10.1086/172580}, \href
  {http://adsabs.harvard.edu/abs/1993ApJ...408..194T} {408, 194}

\bibitem[\protect\citeauthoryear{{Thompson} \& {Duncan}}{{Thompson} \&
  {Duncan}}{2001}]{thompson01}
{Thompson} C.,  {Duncan} R.~C.,  2001, \mn@doi [\apj] {10.1086/323256}, \href
  {http://adsabs.harvard.edu/abs/2001ApJ...561..980T} {561, 980}

\bibitem[\protect\citeauthoryear{{Tiengo} et~al.,}{{Tiengo}
  et~al.}{2013}]{tiengo13}
{Tiengo} A.,  et~al., 2013, \mn@doi [\nat] {10.1038/nature12386}, \href
  {http://adsabs.harvard.edu/abs/2013Natur.500..312T} {500, 312}

\bibitem[\protect\citeauthoryear{{Troja} et~al.,}{{Troja}
  et~al.}{2007}]{Troja07}
{Troja} E.,  et~al., 2007, \mn@doi [\apj] {10.1086/519450}, \href
  {http://adsabs.harvard.edu/abs/2007ApJ...665..599T} {665, 599}

\bibitem[\protect\citeauthoryear{{Troja}, {Piro}, {Vasileiou}, {Omodei},
  {Burgess}, {Cutini}, {Connaughton}  \& {McEnery}}{{Troja}
  et~al.}{2015}]{Troja15}
{Troja} E.,  {Piro} L.,  {Vasileiou} V.,  {Omodei} N.,  {Burgess} J.~M.,
  {Cutini} S.,  {Connaughton} V.,   {McEnery} J.~E.,  2015, \mn@doi [\apj]
  {10.1088/0004-637X/803/1/10}, \href
  {http://adsabs.harvard.edu/abs/2015ApJ...803...10T} {803, 10}

\bibitem[\protect\citeauthoryear{{Troja} et~al.,}{{Troja}
  et~al.}{2017}]{Troja2017}
{Troja} E.,  et~al., 2017, \mn@doi [\nat] {10.1038/nature24290}, \href
  {http://adsabs.harvard.edu/abs/2017Natur.551...71T} {551, 71}

\bibitem[\protect\citeauthoryear{{Troja} et~al.,}{{Troja}
  et~al.}{2018a}]{Troja18b}
{Troja} E.,  et~al., 2018a, preprint, \href
  {http://adsabs.harvard.edu/abs/2018arXiv180806617V} {} (\mn@eprint {arXiv}
  {1808.06617})

\bibitem[\protect\citeauthoryear{{Troja} et~al.,}{{Troja}
  et~al.}{2018b}]{Troja2018}
{Troja} E.,  et~al., 2018b, \mn@doi [\mnras] {10.1093/mnrasl/sly061}, \href
  {http://adsabs.harvard.edu/abs/2018MNRAS.478L..18T} {478, L18}

\bibitem[\protect\citeauthoryear{{Uhm} \& {Zhang}}{{Uhm} \&
  {Zhang}}{2014}]{uhm14}
{Uhm} Z.~L.,  {Zhang} B.,  2014, \mn@doi [\apj] {10.1088/0004-637X/789/1/39},
  \href {http://adsabs.harvard.edu/abs/2014ApJ...789...39U} {789, 39}

\bibitem[\protect\citeauthoryear{{Verbunt}, {Kuiper}, {Belloni}, {Johnston},
  {de Bruyn}, {Hermsen}  \& {van der Klis}}{{Verbunt} et~al.}{1996}]{verbunt96}
{Verbunt} F.,  {Kuiper} L.,  {Belloni} T.,  {Johnston} H.~M.,  {de Bruyn}
  A.~G.,  {Hermsen} W.,   {van der Klis} M.,  1996, \aap, \href
  {http://adsabs.harvard.edu/abs/1996A%26A...311L...9V} {311, L9}

\bibitem[\protect\citeauthoryear{{Xie}, {Zrake}  \& {MacFadyen}}{{Xie}
  et~al.}{2018}]{Xie2018}
{Xie} X.,  {Zrake} J.,   {MacFadyen} A.,  2018, \mn@doi [\apj]
  {10.3847/1538-4357/aacf9c}, \href
  {http://adsabs.harvard.edu/abs/2018ApJ...863...58X} {863, 58}

\bibitem[\protect\citeauthoryear{{Yi}, {Wu}, {Wang}  \& {Dai}}{{Yi}
  et~al.}{2015}]{yi15}
{Yi} S.-X.,  {Wu} X.-F.,  {Wang} F.-Y.,   {Dai} Z.-G.,  2015, \mn@doi [\apj]
  {10.1088/0004-637X/807/1/92}, \href
  {http://adsabs.harvard.edu/abs/2015ApJ...807...92Y} {807, 92}

\bibitem[\protect\citeauthoryear{{Yu}, {Zhang}  \& {Gao}}{{Yu}
  et~al.}{2013}]{yu13}
{Yu} Y.-W.,  {Zhang} B.,   {Gao} H.,  2013, \mn@doi [\apjl]
  {10.1088/2041-8205/776/2/L40}, \href
  {http://adsabs.harvard.edu/abs/2013ApJ...776L..40Y} {776, L40}

\bibitem[\protect\citeauthoryear{{Yu}, {Liu}  \& {Dai}}{{Yu}
  et~al.}{2018}]{yu17}
{Yu} Y.-W.,  {Liu} L.-D.,   {Dai} Z.-G.,  2018, \mn@doi [\apj]
  {10.3847/1538-4357/aac6e5}, \href
  {http://adsabs.harvard.edu/abs/2018ApJ...861..114Y} {861, 114}

\bibitem[\protect\citeauthoryear{{Zhang} \& {M{\'e}sz{\'a}ros}}{{Zhang} \&
  {M{\'e}sz{\'a}ros}}{2001}]{zhang01}
{Zhang} B.,  {M{\'e}sz{\'a}ros} P.,  2001, \mn@doi [\apjl] {10.1086/320255},
  \href {http://adsabs.harvard.edu/abs/2001ApJ...552L..35Z} {552, L35}

\bibitem[\protect\citeauthoryear{{Zhang} \& {Yan}}{{Zhang} \&
  {Yan}}{2011}]{zhangyan11}
{Zhang} B.,  {Yan} H.,  2011, \mn@doi [\apj] {10.1088/0004-637X/726/2/90},
  \href {http://adsabs.harvard.edu/abs/2011ApJ...726...90Z} {726, 90}

\bibitem[\protect\citeauthoryear{{Zhang}, {Fan}, {Dyks}, {Kobayashi},
  {M{\'e}sz{\'a}ros}, {Burrows}, {Nousek}  \& {Gehrels}}{{Zhang}
  et~al.}{2006}]{zhang06}
{Zhang} B.,  {Fan} Y.~Z.,  {Dyks} J.,  {Kobayashi} S.,  {M{\'e}sz{\'a}ros} P.,
  {Burrows} D.~N.,  {Nousek} J.~A.,   {Gehrels} N.,  2006, \mn@doi [\apj]
  {10.1086/500723}, \href {http://adsabs.harvard.edu/abs/2006ApJ...642..354Z}
  {642, 354}

\bibitem[\protect\citeauthoryear{{van Putten} \& {Della Valle}}{{van Putten} \&
  {Della Valle}}{2018}]{vanputten18}
{van Putten} M.~H.~P.~M.,  {Della Valle} M.,  2018, \mn@doi [\mnras]
  {10.1093/mnrasl/sly166}, \href
  {http://adsabs.harvard.edu/abs/2018MNRAS.tmpL.168V} {}

\makeatother
\end{thebibliography}










\bsp	
\label{lastpage}
\end{document}